\documentclass[twocolumn]{bytedance_seed}

\usepackage[toc,page,header]{appendix}

\usepackage{minitoc}

\usepackage[utf8]{inputenc}
\usepackage{textcomp}
\usepackage{array}
\usepackage{amsmath,amssymb}
\usepackage{pifont}
\usepackage{enumitem}
\usepackage{url}
\usepackage{xurl}
\hypersetup{pdftitle={Twindex: Index Cache Offloading with Coarse-to-Fine Selection for Long-Context Sparse-Attention Serving},pdfauthor={Yuxuan Xia, Jiakun Fan, Han Meng, Hao Mark Chen, Shenji Wan, Dongsheng Yang}}
\hypersetup{pdfsubject={},pdfkeywords={}}


\newcommand{\GPU}{\mbox{GPU-A}}
\newcommand{\GPUnode}{\mbox{GPU-A} node}
\newcommand{\GPUB}{\mbox{GPU-B}}

\title{Twindex: Index Cache Offloading with Coarse-to-Fine Selection for Long-Context Sparse-Attention Serving}

\author[1,6\dagger]{Yuxuan Xia}
\author[2,6\dagger]{Jiakun Fan}
\author[3]{Han Meng}
\author[4,6\dagger]{Hao Mark Chen}
\author[1]{Shenji Wan}
\author[5,6]{Dongsheng Yang}

\affiliation[1]{New York University}
\affiliation[2]{Virginia Polytechnic Institute and State University}
\affiliation[3]{University of California, Merced}
\affiliation[4]{Imperial College London}
\affiliation[5]{Princeton University}
\affiliation[6]{ByteDance} \contribution[\dagger]{Work done during internship at ByteDance}

\abstract{
In long-context decoding, GPU memory (HBM) caps how many requests a GPU decodes at once. Sparse-attention models such as Qwen3.8-Flash-Next, DeepSeek V4.1-Flash and GLM-5.2 add an indexer: at every step it scores all past index entries with its own indexer cache, and attention reads the main KV of only the top-k entries. We find that once serving systems move the main KV to host memory, the indexer cache becomes the HBM bottleneck, at 63\% of per-request HBM on Qwen and 75\% on V4.1 at 1M tokens. Therefore, we introduce \textbf{Twindex}, which keeps only a compact copy of the indexer cache in HBM and the full indexer cache in host memory. At each step, the copy selects a larger candidate set, and the model's own indexer picks the top-k from the original rows of the candidates in host memory. The copy combines the shared component with the compressed residuals, both learned once offline. Twindex changes no model weights and runs inside CUDA graphs. At 1M tokens on modern GPUs, Twindex fits \textbf{1.78\ensuremath{\times}}, \textbf{1.62\ensuremath{\times}} and \textbf{3.03\ensuremath{\times}} as many requests as KV offloading alone on Qwen, V4.1 and GLM-5.2, and reaches \textbf{1.44\ensuremath{\times}}, \textbf{1.25\ensuremath{\times}} and \textbf{2.54\ensuremath{\times}} decode throughput when host memory does not limit the batch. Every request keeps its own index, and Qwen and GLM requests share prefix main KV in host memory. In our evaluations, mean benchmark scores are at most 3.1\% below the baseline's.
}

\checkdata[Code]{Pending internal review; will be released soon.}

\begin{document}
\maketitle


\section{Introduction}\label{sec:1}

Large language models increasingly serve long-context workloads: coding agents work over whole repositories, and agent histories grow with every tool call and result \citep{ref46}. At these lengths, decoding is constrained by GPU memory (HBM) capacity and bandwidth. Indexer-based sparse attention \citep{ref4}, used by Qwen3.8-Flash-Next (Qwen), DeepSeek V4.1-Flash (V4.1) and GLM-5.2 (GLM) \citep{ref5,ref6,ref7}, reduces attention work: a lightweight indexer scores entries across the history, and attention reads the key--value (KV) entries corresponding to its top-$k$ selections. HiSparse \citep{ref15} offloads the main KV cache to host memory, retaining only a small buffer in HBM.

The indexer cache, which stores the keys used for selection, remains in HBM. After main-KV offload, it is the largest per-request HBM allocation: at 1M tokens, it accounts for 63\% on Qwen, 75\% on V4.1 and 90\% on GLM, with its share growing with context (\S\ref{sec:2.3}). Its footprint limits GPU batch capacity, while full-history scans consume HBM bandwidth. HiSparse and other offloading systems keep the whole native index in HBM \citep{ref15,ref16,ref17,ref18,ref75}. Methods that shrink the index reuse another layer's top-k \citep{ref9}, score a shared latent index \citep{ref77} or 1-bit keys \citep{ref70}, or train a new indexer that fetches candidates for the native indexer \citep{ref10} (Table~\ref{tab:rw}). The first three replace the model's own top-k with a reused or approximate one, and the last requires training.

The indexer's access pattern makes its cache harder to offload than the main KV cache. Attention reads only selected KV entries, whereas the indexer reads every index key at every step. Offloading the index would transfer about 397\,MiB per Qwen request per step, requiring about 8\,ms of transfer time at our measured 50\,GB/s PCIe bandwidth per GPU. Keeping lower-precision keys in HBM avoids this transfer but can change the selection: an FP4 copy, at about half the original size, recalls only 84.0\% of Qwen's top-$512$ entries at 992K tokens (\S\ref{sec:2.4}).

We present Twindex, which separates candidate screening from final selection. It keeps compressed keys in HBM and the model's original index keys (native rows) in host memory. Each step scans the compressed keys to form a larger candidate set, then rescores those candidates using their native rows and the model's own indexer. This recovers the native top-$k$ whenever the candidates contain it. The model's own indexer selects the final top-$k$ within the candidate set, without retraining. A small HBM buffer caches recently used native rows, so only candidate misses cross PCIe. To make the keys compact, Twindex fits a position-dependent component shared across inputs offline and compresses only the residual, using calibration queries to limit score error. Like SAKI \citep{ref71}, Twindex subtracts a calibrated mean before rotation and fits the compression in the query metric, and it adds native rescoring and an index hot buffer for the indexer cache. On held-out Qwen contexts, 36\,B keys (about a quarter of a native row) and a screening budget of $M=1{,}280$ yield candidates covering 99.6\% of the attention mass assigned to the native top-$512$ entries (\S\ref{sec:3.3}). This design reduces HBM use and scan traffic and limits the cost of fetching native rows from host memory.

We implement Twindex in SGLang \citep{ref41} on top of HiSparse, without changing model weights or attention kernels, and compare it with HiSparse-Opt, which includes the same serving optimizations (\S\ref{sec:5.2}). On modern GPUs at 1M tokens, Twindex serves 1.78\ensuremath{\times}, 1.62\ensuremath{\times} and 3.03\ensuremath{\times} as many concurrent requests on Qwen, V4.1 and GLM, respectively, and achieves 1.44\ensuremath{\times}, 1.25\ensuremath{\times} and 2.54\ensuremath{\times} the decode throughput. Across our evaluations on all three models, any observed decrease in mean benchmark score is at most 3.1\% relative to the baseline. V4.1 keeps every request's prefix state private, so its gain needs no shared prefix. On Qwen and GLM, requests share main-KV prefixes in host memory (\S\ref{sec:5.1}), because our test nodes have no cache tier below host memory. Production serving stacks keep KV caches in such tiers, on local SSD or in remote pools \citep{ref25,ref44,ref79,ref80,ref81}; once a tier holds most of the main KV, both systems run at their HBM limits, where Twindex admits 1.78\ensuremath{\times} and 3.03\ensuremath{\times} as many Qwen and GLM requests (\S\ref{sec:5.3}).

\begin{table}[t]
\centering
\footnotesize
\setlength{\tabcolsep}{3pt}
\caption{Comparison of index-cache designs.}\label{tab:rw}
\begin{tabular}{@{}lllll@{}}
\toprule
Design & HBM key & Layers & Native rows & Top-k \\
\midrule
HiSparse v1 & native & all & HBM & native \\
HiSparse v2 & shared & all & HBM & native \\
IndexCache & native & 1/4 & dropped & reused \\
LatentIndex & latent & grouped & dropped & latent \\
Self-Indexing & 1-bit & all & dropped & 1-bit \\
FlashMemory & native & 3/21 & host & native$^\dagger$ \\
Twindex & compressed & all & host & native on M \\
\bottomrule
\end{tabular}

\vspace{1pt}{\scriptsize $^\dagger$FlashMemory uses a trained screen before native selection within the fetched subset.\par}
\end{table}

This paper makes three contributions:
\begin{enumerate}
\item We reveal that \textbf{the indexer cache is the HBM bottleneck, and host memory can become the next limit.} We measure per-request HBM after main-KV offload on three indexer designs at up to 1M tokens. The indexer cache is the largest item on all three models, and its share grows with context. Offloading it is too slow, and quantizing it changes the model's selection (\S\ref{sec:2}). Once it is compressed, host memory can limit the decode batch before HBM does (\S\ref{sec:5.3}).
\item \textbf{Twindex.} Twindex screens with short compressed residuals in HBM and rescores the candidates with native rows from host memory, so the model's own indexer picks the final top-k from the candidates without training. One offline fit captures the shared component that keeps the compressed keys short (\S\ref{sec:3}, \S\ref{sec:4}).
\item \textbf{An evaluation against HiSparse with the same serving optimizations}, whose generic fixes we also contribute, on three models at 1M tokens (\S\ref{sec:5}).
\end{enumerate}

\begin{figure}[tp]
\centering
\includegraphics[width=\columnwidth]{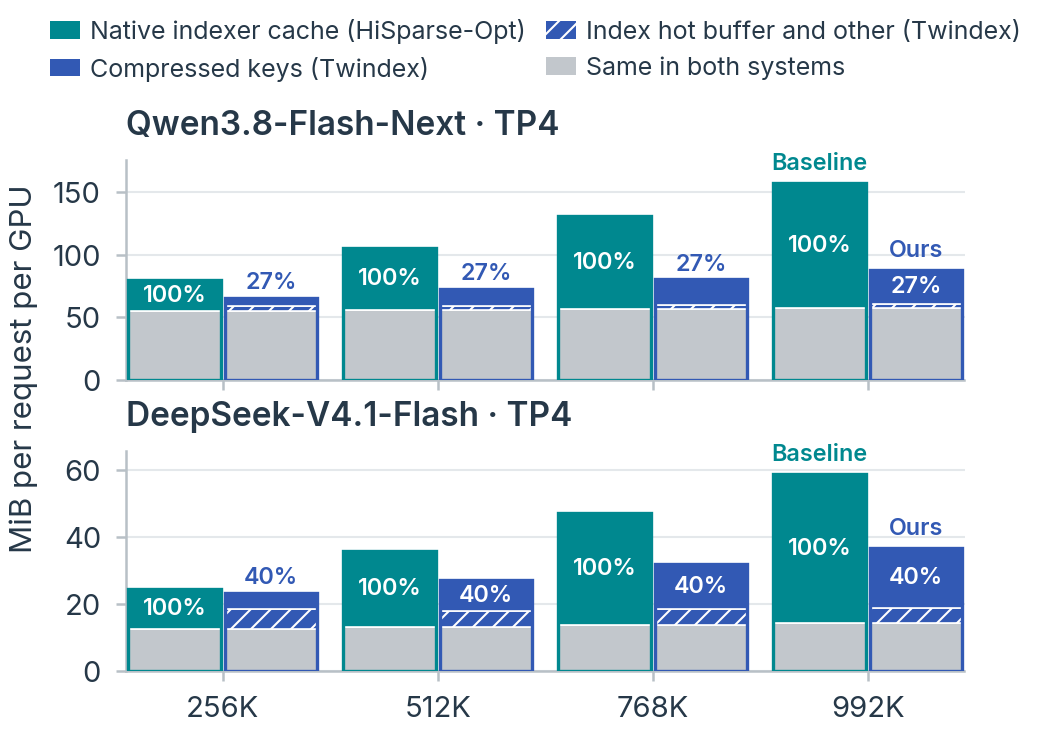}
\caption{The index is the largest per-request item in HBM. Twindex shrinks the index by a fixed ratio. Each pair shows the per-request HBM per GPU of HiSparse-Opt (left) and Twindex (right) in the serving run at that length, at the full-HBM limit (\S\ref{sec:2.3}). Percentages give each index's size relative to HiSparse-Opt's native index. Twindex's index hot buffer has a fixed size, and on V4.1 the hatched segment also holds Twindex's other memory above HiSparse-Opt's.}\label{fig:7}
\end{figure}

\section{Background \& Motivation}\label{sec:2}

\subsection{Indexer-Based Sparse Attention}\label{sec:2.1}

Indexer-based sparse attention first selects k past index entries and then attends to their main-KV entries. Each request therefore keeps two caches. The main KV cache holds the keys and values that attention reads, and the indexer cache holds a small index key for every past position. At each decode step, a lightweight indexer maps the current token to a few index queries $q_h$ and scores every past position t as $s_t = \sum_h w_h\,\sigma(q_h\cdot k_t)$, where $k_t$ is the index key, $\sigma$ a fixed nonlinearity and $w_h$ per-head weights. The model keeps the k highest-scoring positions, its top-k, and attends only to their main-KV entries.

DeepSeek Sparse Attention (DSA) \citep{ref4} introduced this design. An index entry is one row of one indexer layer. Qwen \citep{ref5} stores one FP8 entry of 132 B per 4-token block in 12 indexer layers, interleaved with linear-attention layers \citep{ref12}. V4.1 \citep{ref6} stores FP4 entries of 68 B in four indexer layers, and later layers select from the blocks that layer 20 picks. GLM \citep{ref7} stores one FP8 entry of 132 B per token in each of 21 indexer layers. Sparse access saves computation but not memory: every past main-KV entry and index key stays in memory, so at one million tokens memory limits how many requests a GPU can serve.

\subsection{Serving Long-Context DSA with HiSparse}\label{sec:2.2}

HiSparse \citep{ref15} moves the main KV cache to host memory (Figure~\ref{fig:1}a). Each request keeps its full main KV in host memory and a main-KV buffer of recently used entries in HBM. Only the selected entries that are missing from the buffer are copied in over the host link. The copies are small, because consecutive steps pick mostly the same entries in our measurements, and HiSparse-Opt runs them while the GPU computes the rest of the step (\S\ref{sec:5.2}). On V4.1 at 1M tokens, they make a 45 ms step only 1.3 ms longer. The main KV that a request holds in each GPU's HBM drops from 12 GiB to a 24 MiB buffer on Qwen and from 720 MiB to 4.7 MiB on V4.1. The indexer already names the entries to fetch, so no prediction or search is needed.

\subsection{The Remaining Indexer-Cache Bottleneck}\label{sec:2.3}

After HiSparse moves the main KV to host memory, the indexer cache is the largest per-request item in HBM, and it sets how many requests fit on a GPU. Two per-request caches grow linearly with the context, the main KV and the indexer cache, and the rest of a request's state has a fixed size. HiSparse leaves the indexer cache in HBM because the indexer reads every index key at every step.

Prior work has seen parts of this problem. ESS \citep{ref17} measures the indexer cache at 16.8\% of all cache storage. Self-Indexing \citep{ref70} reports a capacity gain from a smaller index on one model at about 128K tokens, and FlashMemory-DeepSeek-V4 \citep{ref10} offloads most index keys of one model. We measure the indexer cache's share of per-request HBM after main-KV offload in HiSparse-Opt, our baseline (\S\ref{sec:5.2}), on three models whose indexers differ in precision, heads and unit of selection, at 256K to 1M tokens and at the full-HBM limit of 178 of 179 GiB per GPU. At 992K tokens, the indexer cache takes 63\% of per-request HBM on Qwen and 75\% on V4.1, although V4.1's index is already FP4 (Figure~\ref{fig:7}), and 90\% on GLM-5.2 (336 of 374 MiB). Since only the index grows with context, its share rises from 31\% at 256K to 63\% at 1M tokens on Qwen and from 49\% to 75\% on V4.1. The indexer cache caps the decode batch on every design we test, and the cap tightens with context.

\subsection{Opportunities for Compact Indexing}\label{sec:2.4}

\textbf{Observation 1: native rescoring improves selection recall.} The indexer's scores only choose an unordered set of positions, so a score change that moves no entry across the top-k boundary leaves the selection unchanged. We measure a selection by its recall, the share of the native top-512 entries it contains. Selecting the top-512 from an FP4 copy of Qwen's index, 68 B instead of 132 B per row, recalls 84.0\% at 992K tokens. Rescoring the copy's top 1,280 entries with their native rows raises recall to 99.6\%. Fewer bits alone do not pay off on \GPU{}, our evaluation GPU (\S\ref{sec:5.1}), which lacks 2- and 3-bit Tensor Core multiplies \citep{ref62}, so a smaller copy must also keep fewer dimensions.

\begin{figure*}[tp]
\centering
\includegraphics[width=\textwidth]{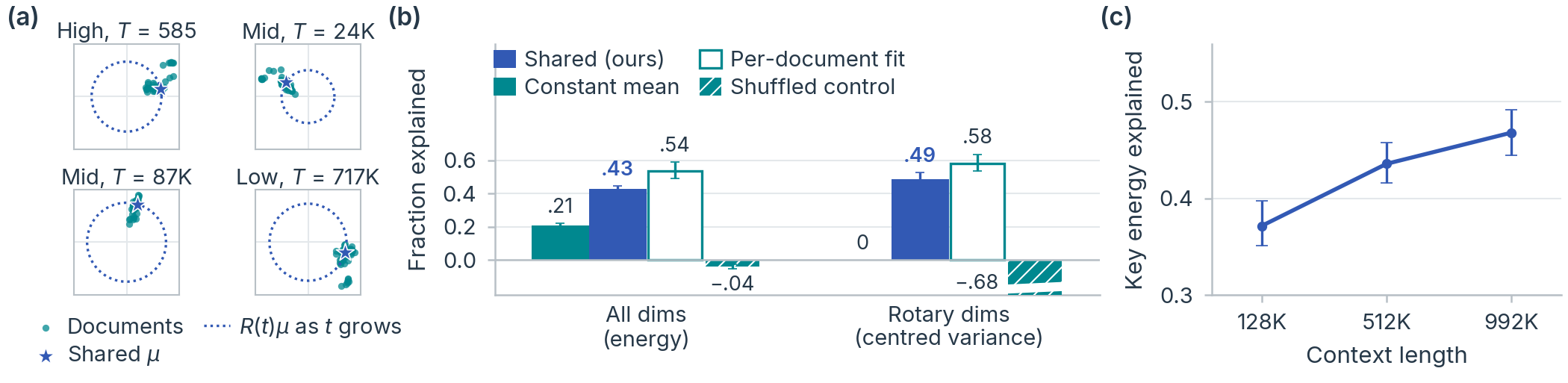}
\caption{Index keys share a component that varies with position and is common across inputs (Qwen). (a) One rotary pair per panel, high to low frequency, with period T tokens. Dots are per-document mean keys before rotation, the star is the shared mean $\mu$, and the dotted circle is the shared component $R(t)\,\mu$, which turns once every T tokens. (b) Energy explained by each predictor. (c) By context length.}\label{fig:3}
\end{figure*}

\textbf{Observation 2: removing a shared component improves recall at a fixed compressed-key size.} With rotary position embedding (RoPE) \citep{ref58}, the index key at position $t$ is $k_t = R(t)\,u_t$, where $u_t$ is the key before rotation. If $u_t$ has a mean $\mu$ shared across inputs, every key contains $R(t)\,\mu$, a component set by position alone, which we call the shared component. SAKI \citep{ref71} also subtracts a mean before rotation when it compresses dense attention keys, but it does not test whether this mean is shared across inputs. On held-out Qwen keys from 10 domains at 128K--992K tokens, the per-document means lie close to one $\mu$ (Figure~\ref{fig:3}a), and one $\mu$ fitted on four calibration documents removes 43\% of the keys' summed squared norm, against 54\% for a fit to each document, 21\% for a constant mean and none for $R(t)\,\mu$ at shuffled positions (Figure~\ref{fig:3}b). The share grows with context (Figure~\ref{fig:3}c) and holds in every indexer layer and domain (Appendix~\ref{app:B}.1).

\begin{figure}[tp]
\centering
\includegraphics[width=\columnwidth]{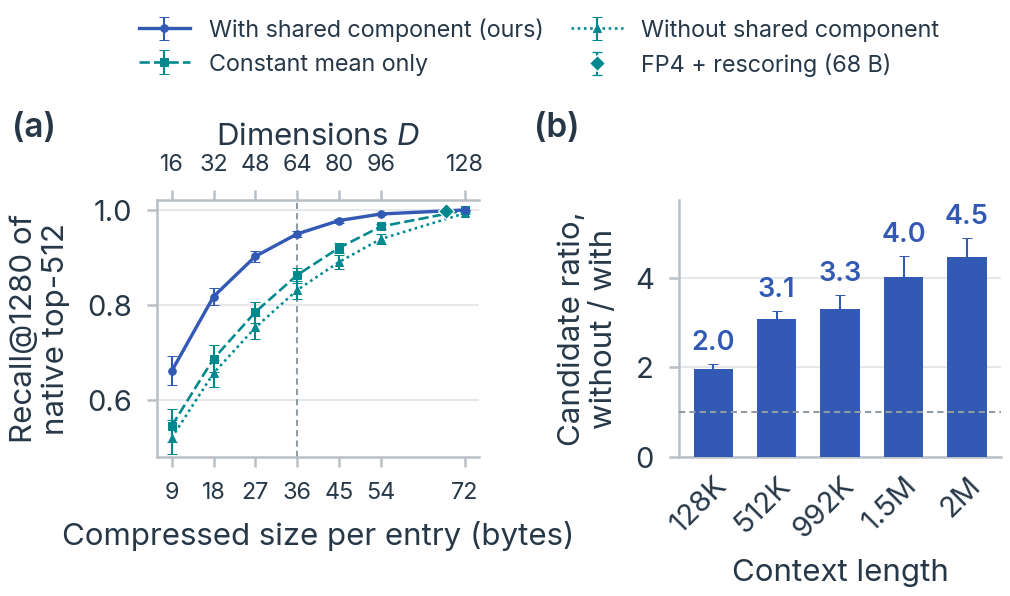}
\caption{Recall with and without the shared component (Qwen, held-out documents, NVFP4). The dashed line marks the deployed 36 B compressed key. (a) Recall@1280 against compressed size. (b) Candidates needed for 99.5\% coverage without the shared component, divided by those needed with it, at 36 B and at each context length we measured.}\label{fig:4}
\end{figure}

Removing this component raises recall at a fixed compressed-key size. If the shared part stays in the keys, the compression spends two of its $D$ dimensions on this rotating mean in each rotary pair that the mean dominates. At 36 B per entry (64 dimensions in FP4), the 1,280 entries that the compressed residual ranks highest contain 94.9\% of the native top-512 entries (the recall), against 83.1\% when the compression is fitted to the keys themselves (Figure~\ref{fig:4}a). Covering 99.5\% of the native top-512 takes about 2.4K candidates with the shared component removed and 6.7K without, and the gap grows with context (Figure~\ref{fig:4}b). V4.1 and GLM show the same, with recall .994 against .899 and .997 against .710 (Table~\ref{tab:abl}).

\section{Twindex Design}\label{sec:3}

\begin{figure*}[tp]
\centering
\includegraphics[width=\textwidth]{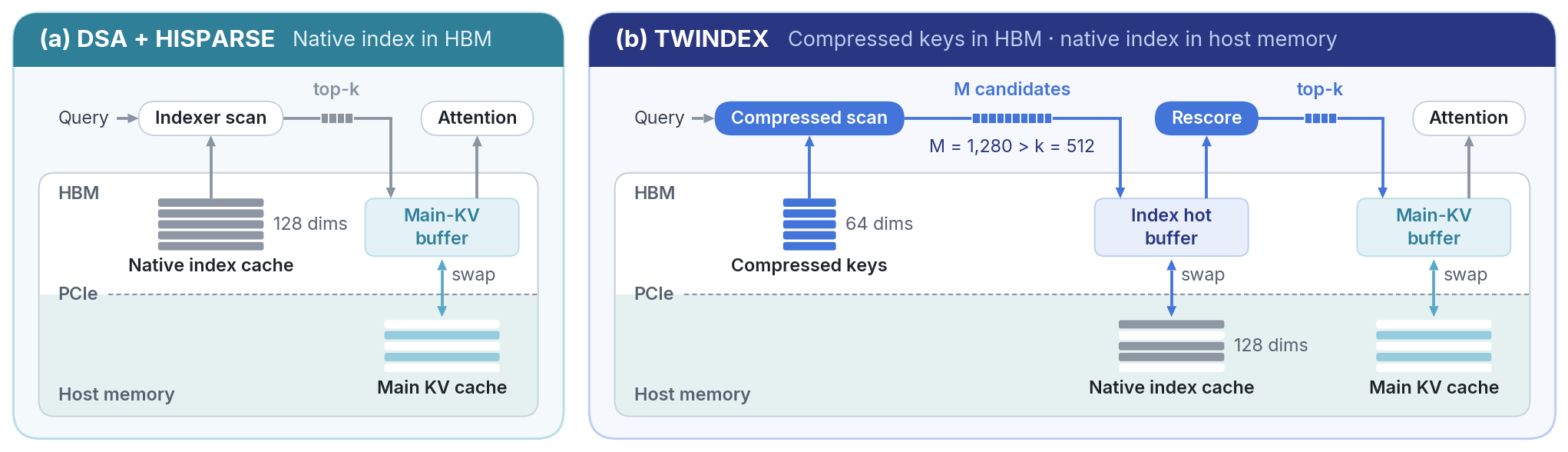}
\caption{One decode step of (a) DSA with HiSparse and (b) Twindex, in the Qwen setting. Each row is one index entry. A native index row is the key that the model's own indexer stores, and a compressed key is a 36 B summary of it. The candidates are the M entries with the highest compressed scores.}\label{fig:1}
\end{figure*}

Twindex splits the indexer cache across HBM and host memory, as HiSparse does for the main KV (Figure~\ref{fig:1}b). HBM holds a short compressed key of every index entry and an index hot buffer of recently used native rows. Host memory holds every native row. Each decode step scans the compressed keys of the whole history and keeps M candidates, with M larger than the number k of selected entries. Rescoring then reads the candidates' native rows through the index hot buffer, scores them with the model's own indexer and keeps the top-k, which then selects the main-KV rows to read as in HiSparse.

\subsection{Residual Index Compression}\label{sec:3.2}

\textbf{Compress the residual, not the key.} Twindex replaces each native index entry in HBM with a short compressed key built for one purpose: to rank entries the way the native scores do. For each indexer layer, Twindex stores the calibration mean of the keys before rotation, one $d$-dimensional vector, where $d$ is the dimension of a native index key. On Qwen, $d = 128$, and RoPE covers 64 of these dimensions, a count unrelated to the compressed width D below. The shared component of the entry at position p is $m(p) = [R(p)\,v;\ \mu_n]$, where v, the rotary half of the mean, is rotated by the model's own RoPE at p, and $\mu_n$ is the mean of the non-rotary half. Twindex compresses only the residual $r_p = k_p - m(p)$. As $q\cdot k_p = q\cdot m(p) + q\cdot r_p$ and the first term is computed exactly from the position, the compressed key only has to preserve $q\cdot r_p$. Twindex fits all parameters of the compression once per model, offline, on calibration documents (\S\ref{sec:adapt}).

\textbf{Projection for ranking, not reconstruction.} A linear compression of width $D < d$, the number of dimensions that each compressed key keeps, stores $z_p = r_p A$ for each entry and maps the query once per step to $qB$, with $A, B \in \mathbb{R}^{d\times D}$. The residual score is approximated by $(qB)\cdot z_p$. Standard PCA \citep{ref67} chooses A to minimize the reconstruction error $\mathbb{E}\lVert r-\hat r\rVert^2$, which treats every direction of the key as equally important. The indexer sees a key only through its scores, so the error that matters is the score error. For queries with second-moment matrix $\Sigma_q = \mathbb{E}[qq^\top]$,
\[
\mathbb{E}_q\big[(q\cdot r - q\cdot \hat r)^2\big] = (r-\hat r)^\top\, \Sigma_q\, (r-\hat r),
\]
the reconstruction error measured in the metric $\Sigma_q$. Twindex therefore runs PCA in the query metric. For a metric G, let $H = G^{1/2}$ and let $U_D$ hold the top-D eigenvectors of the covariance of $rH$ over calibration residuals. Twindex sets $A = H\,U_D$ and $B = H^{-1}U_D$. At $D = d$ the score is exact:
\[
(qB)\cdot(rA) = q\,H^{-1}UU^\top H\,r^\top = q\cdot r .
\]
For $D < d$ and queries independent of the residuals, the compression fitted in the metric $G = \Sigma_q$ is the rank-$D$ linear compression with the smallest expected squared score error on the centred residuals before quantization (Appendix~\ref{app:B}.2). This is a weighted Eckart--Young bound \citep{ref60}. SAKI \citep{ref71} derives the same optimum for attention keys and uses the query metric alone. Twindex fits the metric once per model, offline, from calibration queries, and serves every request with the same compression. Two steps help this one fit transfer to other documents and positions. First, Twindex blends the query metric with the identity, $G = (1-\alpha)\,\Sigma_q/s + \alpha I$, where $s$ is the mean diagonal of $\Sigma_q$. Few calibration queries give almost no weight to directions that other documents use, and the identity keeps a weight of at least $\alpha$ on every direction. Second, Twindex averages the metric and the residual samples over shifts of the positions, so the fit also holds at positions that the calibration tokens did not occupy. The deployed compression adds these two steps and FP4 quantization, so we measure its ranking directly (\S\ref{sec:5.4}).

\textbf{Precision and scoring.} Twindex stores $z_p$ at low precision in a format that the scan's Tensor Cores multiply natively, chosen from the native index's precision (\S\ref{sec:4}). Each step it maps the query once and scores every entry as $\hat s_p = \sum_h w_h\,\sigma\big(q_h\cdot m(p) + (q_hB)\cdot \hat z_p\big)$, with the model's own nonlinearity $\sigma$ and head weights $w_h$, so the compressed scores are on the native score scale.

\subsection{Rescoring Candidates with Native Rows}\label{sec:3.3}

\textbf{The compressed keys pick candidates, and the native rows pick the top-k.} At each decode step Twindex takes the M entries with the highest compressed scores, together with the set $\mathcal{A}$ of entries that always enter the candidate set (the first entries, which act as attention sinks, and the most recent ones). It then reads the native index rows of these candidates through the index hot buffer (\S\ref{sec:3.4}), rescores them with the model's own scorer, and applies the model's own top-k: $\mathcal{C} = \mathcal{A} \cup \mathrm{Top}_M(\hat s)$ and $\mathcal{S} = \mathrm{Top}_k\{ s_j : j \in \mathcal{C} \}$. This describes one candidate pool, and V4.1's layer 20 and GLM build grouped candidate sets. $\mathcal{S}$ equals the native top-k, up to ties, whenever $\mathcal{C}$ contains every native top-k entry: the compressed scores only have to place these entries among the top M, not rank them correctly. Screening can miss a native top-k entry, and $\mathcal{S}$ then holds the next-best candidate in its place. On Qwen, $\mathcal{C}$ holds 99.6\% of the native top-512's attention mass (Appendix~\ref{app:extra}.3).

\subsection{Index Hot Buffer}\label{sec:3.4}

Reading M native rows per layer and request from host memory at every step would put the host link on the critical path. About half of each step's candidates were also candidates in the previous step, so Twindex keeps an index hot buffer in HBM. It is a small per-request, per-layer store of recently read native rows, separate from HiSparse's main-KV buffer. Rescoring reads a candidate's row from the buffer when it is there. Otherwise the buffer swaps it in from host memory and evicts a row that was not used recently. The compressed keys, the index hot buffer and the host rows thus form a memory hierarchy for the indexer cache.

\section{Implementation}\label{sec:4}

We implement Twindex inside SGLang \citep{ref41} on top of HiSparse v1. Twindex replaces the indexer cache, index scan and top-k path, and leaves the model weights, attention kernels and HiSparse's main-KV path unchanged.

\textbf{Compressed-key formats.} Twindex stores compressed keys in NVFP4 \citep{ref62}, the 4-bit format that \GPU{} Tensor Cores multiply natively (E2M1 values with one 8-bit scale per 16 dimensions): 36 B per Qwen entry at $D = 64$ instead of 132 B, and 24 B per GLM entry on average, with $D = 32$, 48 or 64 per layer. V4.1's native index is already FP4 (68 B per entry), so V4.1 stores its leading 16--32 coordinates in FP4 and the next 48 in 2 bits, 24--33 B per entry. Its 2-bit levels are exact FP4 values, so the scan stays on the FP4 Tensor Core path, and it keeps 64 or 80 dimensions. The scan's speed relies on these block-scaled FP4 multiplies, so we measure throughput only on \GPU{}.

\textbf{Index hot buffers.} Qwen uses a ring of 8,192 rows per layer with CLOCK replacement, 3.1 MiB per request per GPU. V4.1 uses a 4-way set-associative buffer with LRU replacement, which serves .68--.87 of the candidates. GLM keeps 2,048--16,384 rows per layer.

\textbf{Kernels.} For each tile of entries, one block-scaled FP4 instruction \citep{ref62} computes the compressed term and applies its scales inside the multiply, and a second adds the shared-component term into the same accumulator. Twindex compresses keys with Triton \citep{ref64} kernels that run in one launch at every index-key store. GLM compresses each 64-token page once the page is complete and scores the newer rows exactly. Rescoring reads the rows that miss the index hot buffer from host memory with a GPU kernel through managed memory, in chunks whose transfer overlaps the next chunk's scan.

\textbf{CUDA graphs.} Every Twindex kernel in the decode step has fixed launch shapes and needs no host synchronization, so we capture the decode step as one CUDA graph \citep{ref63} at batch size C.

\section{Evaluation}\label{sec:5}

\subsection{Experimental Setup}\label{sec:5.1}

\textbf{Hardware and software.} Unless noted, experiments run on \GPUnode{}s. Each node has eight GPUs, two Intel Xeon Platinum 8592+ CPUs with 64 cores each, and 4 TiB of host memory, of which a job may use 3,700 GiB. Each \GPU{} has 179 GiB of HBM and Tensor Cores that multiply NVFP4 matrices natively. Qwen's quality runs use \GPUB{} nodes, whose GPUs have 80 GB of HBM and FP8 but no FP4 Tensor Cores. Twindex's selection does not depend on FP4 Tensor Cores: on \GPUB{}, a Triton kernel scans the same compressed keys with the same D and M, and on 1,860 held-out (layer, step) samples it selects the same final top-512 as the \GPU{} scan. Each GPU reaches host memory over its own PCIe Gen5 x16 link, at 55 GB/s alone and about 50 GB/s when all eight GPUs copy at once. Qwen and V4.1 run with tensor parallelism over four GPUs (TP4) and GLM with TP8, and no other job uses the node's GPUs during a timed run. Most runs measure the decode node of a prefill/decode-disaggregated deployment \citep{ref42,ref43,ref44}.

\textbf{Workload.} Every request is different. Qwen and GLM serve a prefix-sharing workload, and V4.1 shares no prefix state. In the prefix-sharing workload, each 992K-token prompt is one of three 976K-token LongBench v2 \citep{ref51} documents followed by a 16K-token tail unique to the request. V4.1 prompts have the same form, but each request keeps a private copy of its whole prefix state. Every request decodes up to 32,768 tokens greedily.

\textbf{Prefix sharing and host memory.} Which memory limits the batch first depends on the size of the main KV. V4.1's main KV is small, so private copies fit in host memory and HBM limits both systems. Qwen's and GLM's main KVs are large. A 1M-token Qwen request would hold about 47 GiB of it in host memory, and a node would fit fewer than 80 such requests in either system. Our test nodes have no cache tier below host memory, so for these two models requests share prefixes, and both systems store each document's main KV once in host memory, as prefix-reuse systems and APIs do \citep{ref25,ref26,ref41,ref44,ref72,ref73}. The index stays private per request in both systems, because HiSparse v1 does not share index pages. Twindex's native index adds 0.37 GiB of host memory per Qwen request.

\textbf{Metrics.} Throughput \textbf{Q} is decode tokens per second, $Q = C/E(C)$, where \textbf{C} is the number of running requests and \textbf{E(C)} the mean decode-step time over a fixed steady-state window. The throughput ratio \textbf{R} $= Q_{\text{Twindex}} / Q_{\text{HiSparse-Opt}}$ compares the two systems on the same node, GPUs and snapshots. Latency is the median and p95 time per output token (\textbf{TPOT}) and the time to first token (\textbf{TTFT}).

\subsection{Adapting Twindex to Each Model}\label{sec:adapt}

\textbf{Requirements.} Twindex runs the same algorithm (\S\ref{sec:3}) on every model. It needs an indexer that scores entries as $\sum_h w_h\,\sigma(q_h\cdot k)$, an engine that exposes each index key when it stores it, and a top-k that follows the scores. The shared component uses the RoPE part of each key, and a constant mean covers the rest.

\textbf{Calibration.} Twindex fits its compression once per model on public calibration documents. Each capture records the index keys of one document prefix and the indexer queries of a few decode steps after it. Qwen uses five captures of four documents at 512K and 1M tokens, V4.1 one 512K-token capture of each of these documents, and GLM one capture of each of 18 other documents. Each fit averages over position shifts of up to $\pm$262,144. Except for the explicitly labeled per-document fits, every reported recall and attention mass comes from documents outside the fit that produces it (Appendix~\ref{app:B}.5): 42 Qwen captures from 10 domains at 128K--2M tokens, six 992K-token GLM documents and six 512K-token V4.1 documents, where each of V4.1's four calibration documents is scored by a fit on the other three.

\textbf{Choosing D, M and $\alpha$.} We select D, M, and $\alpha$ on development documents disjoint from both the fitting data and the final evaluation documents, and freeze them before the main serving and task-quality evaluations. D sets the bytes per entry and the scan time, M sets how many native rows each step fetches and rescores, and $\alpha$ sets how much the fit relies on the calibration queries. On each model, we fit the compression at D from 16 to 128 on Qwen, 48 to 128 on V4.1 and 32 to 64 on GLM, and at $\alpha$ from 0 to 1, measure recall against M, and choose by hand the smallest compressed key whose M reaches high recall at an affordable fetch cost, checked with paired throughput runs. GLM sets each layer's $M_l$ by a fixed rule: the M that reaches recall .98 on 99\% of the steps of development documents, raised in steps of 1,024 until the mean recall at 992K is at least .9979. Qwen and V4.1 use $\alpha = 0.5$, and GLM uses $\alpha = 0$. The sensitivity studies vary one of them at a time (\S\ref{sec:sens}).

\textbf{Model-specific engineering.} Model-specific choices include the index layout, candidate granularity, calibration weighting and quantization-scale fitting (Appendix~\ref{app:C} and Table~\ref{tab:1} in Appendix~\ref{app:D}). Twindex selects 4-token blocks on Qwen, one or two tokens on V4.1 and single tokens on GLM. At V4.1's layer 20, one fetch of 6,144 candidate blocks serves both its own top-512 and the 2,048 blocks from which later layers select. On GLM, each GPU rescores candidates among its own index pages, and the per-GPU results merge into the model's native top-2,048.

\subsection{Baselines}\label{sec:5.2}

Our baseline, \emph{HiSparse-Opt}, is HiSparse v1 with serving optimizations that both systems use (Appendix~\ref{app:A}), and an optimization counts toward Twindex only if R grows after both systems adopt it. Two of them fix problems in SGLang v0.5.22, and we contribute both. HiSparse-Opt stores each request's index once across the TP ranks instead of on every rank and fetches each missed main-KV row once per KV group, which alone makes the V4.1 baseline up to 1.72\ensuremath{\times} faster. It also scores Qwen's FP8 index with DeepGEMM \citep{ref61} at 6.5 instead of 2.2 TB/s.

\textbf{Top-k exactness problem in SGLang.} SGLang's top-k kernel for Qwen can return a top-k that misses higher-scoring entries at long contexts. It groups the indexer's scores into bins and copies the bin that holds the k-th score into a 4,096-entry buffer, which it sorts exactly. When that bin holds more entries, the kernel drops the rest without an error. The overflow appears from 512K tokens on. At 992K, an average of 4.3 of the 512 returned entries (0.85\%) are not in the exact top-512, and 93\% of these rank 513--640. HiSparse-Opt keeps this kernel as SGLang ships it. On Qwen, Twindex ranks at most 1,408 candidates, which always fit the buffer, so its final top-k is exact over its candidates. V4.1's top-k kernel returns the exact top-512 at 512K and 1M tokens, and GLM's top-k uses an exact sort. All selection metrics in this paper use the exact top-k of the native scores. 

\subsection{Performance}\label{sec:5.3}

\emph{Throughput at the full-HBM limit.} Twindex trades a longer step for more requests per step. Each system runs at the concurrency with its highest measured throughput under the full-HBM limit (Figure~\ref{fig:tput}, Table~\ref{tab:2} in Appendix~\ref{app:extra}). On Qwen, Twindex serves 1.78\ensuremath{\times} as many requests with a 1.24\ensuremath{\times} longer step, which gives 1.44\ensuremath{\times} the throughput (R = 1.438). R is 1.251 on V4.1 and 2.540 on GLM. On Qwen, the gain holds over long runs (R is 1.47 over the first 8K steps).

\begin{figure}[tp]
\centering
\includegraphics[width=\columnwidth]{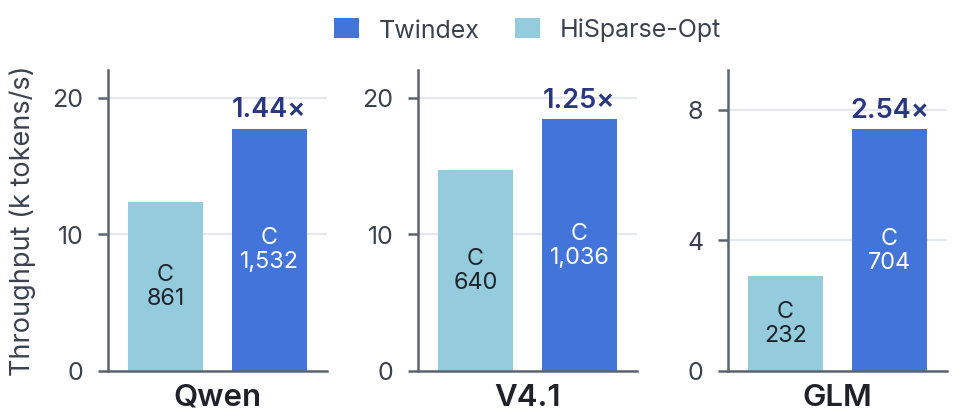}
\caption{Decode throughput at 992K tokens. Each system runs at the concurrency C with its highest throughput under the full-HBM limit. C is inside each bar, and R is above each Twindex bar.}\label{fig:tput}
\end{figure}

\emph{Throughput by context length.} R rises with context (Table~\ref{tab:3}). A longer context gives the index a larger share of each request's HBM and gives compression more to save. From 256K to 1M, R rises from 1.06 to 1.44 on Qwen and from 0.95 to 1.25 on V4.1. On Qwen at 256K, host memory, not HBM, caps Twindex's concurrency. With many distinct prefixes in a Qwen batch, R is 1.46 with three, 1.44 with eight and 1.42 with 16 (Figure~\ref{fig:11} in Appendix~\ref{app:extra}).

\emph{Prefix share and cache tiers.} Our 992K-token Qwen and GLM prompts share 98\% of their tokens, a share we choose to fit our test nodes' host memory. In a more general setting that shares no prefix, the private main KV belongs in a cache tier below host memory. Production serving stacks such as SGLang HiCache, Mooncake, Dynamo, LMCache and CachedAttention keep KV caches in tiers below host memory, on local SSD or in remote pools reached over RDMA \citep{ref25,ref44,ref79,ref80,ref81}. Most of the cache can stay there: a request's selections touch only 13--19\% of its context blocks over 8K decode steps in our traces, so a small per-request buffer in host memory cuts reads from the tier. Both systems then run at their HBM limits, where Twindex admits 1.78\ensuremath{\times} and 3.03\ensuremath{\times} as many Qwen and GLM requests.

\begin{table}[tp]
\centering
\footnotesize
\setlength{\tabcolsep}{4pt}
\caption{Throughput ratio R by context length. C is the number of concurrent requests in HiSparse-Opt and in Twindex. $^\dagger$Host memory, not HBM, limits Twindex's C in this cell.}\label{tab:3}
\begin{tabular}{@{}lrrrr@{}}
\toprule
 & \multicolumn{2}{c}{Qwen} & \multicolumn{2}{c}{V4.1} \\
\cmidrule(lr){2-3}\cmidrule(lr){4-5}
Context & C & R & C & R \\
\midrule
256K & 1,649 / 1,943$^\dagger$ & 1.057 & 1,520 / 1,600 & 0.947 \\
512K & 1,265 / 1,836 & 1.163 & 1,162 / 1,584 & 1.016 \\
768K & 1,026 / 1,677 & 1.297 & 879 / 1,340 & 1.089 \\
992K & 861 / 1,532 & 1.438 & 640 / 1,036 & 1.251 \\
\bottomrule
\end{tabular}
\end{table}

\emph{Steady-state decode under prefill/decode deployment.} The decode-node runs give nearly the same steady-state decode throughput as a prefill/decode deployment (Table~\ref{tab:2} in Appendix~\ref{app:extra}), in which one prefill and one decode instance share a node with four GPUs each and Mooncake \citep{ref44} moves each request's state between them. On V4.1, the prefill instance also compresses every request's index. On Qwen, the prefill instance prefills only the last 64 requests, because 1,532 fresh 1M-token prefills would take over 12 hours, and the decode instance restores the rest from snapshots. Each system stays within 3.3\% of its decode-node run on the same node, and R is 1.412 against 1.463 on Qwen and 1.225 on V4.1.

\begin{figure}[tp]
\centering
\includegraphics[width=\columnwidth]{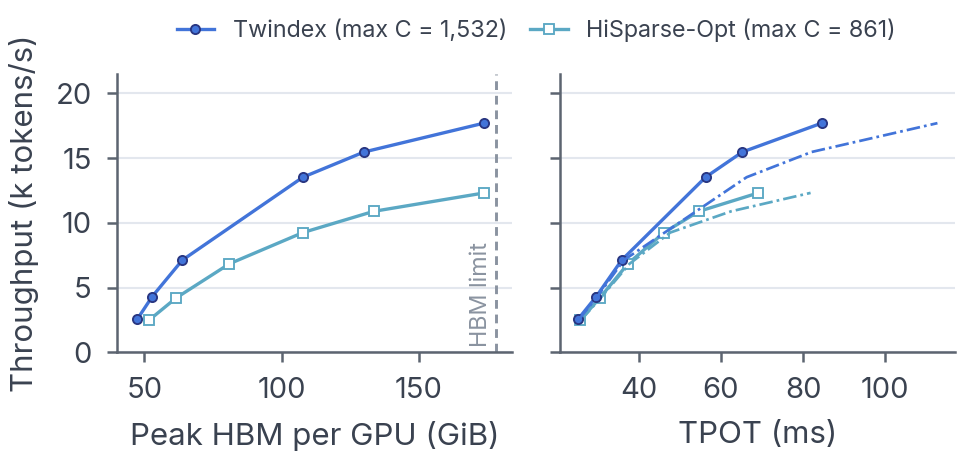}
\caption{Throughput by HBM per GPU (left) and by TPOT limit (right), Qwen. Each marker is one measured concurrency, from C = 64 to each system's full-HBM limit. Right: median TPOT (solid) and p95 (dash-dot).}\label{fig:6}
\end{figure}

\emph{Throughput with less HBM.} Twindex's gain grows when each GPU has less HBM for requests (Figure~\ref{fig:6}, left). On Qwen, R is 1.73 with about 52 GiB of HBM per GPU and 1.46 with 108 GiB. On V4.1, caps of 170, 160 and 150 GiB per GPU raise R to 1.26, 1.28 and 1.32.

\setlength{\parskip}{6pt plus 3pt}
\subsection{Latency and Step-Time Breakdown}\label{sec:lat}

\emph{TPOT at equal concurrency.} At equal concurrency C, Twindex's TPOT, the time between a request's consecutive output tokens, is lower than HiSparse-Opt's on Qwen and GLM (Table~\ref{tab:lat}). Each request adds less to Twindex's step because the scan of compressed keys reads 36 B per entry on Qwen and 24 B on GLM instead of 132 B. On Qwen, the scan costs 5.7 \textmu{}s per request per step against 15.0 \textmu{}s for the native indexer. On V4.1, whose native index is already FP4, Twindex's step is about 4--7\% longer at equal C, so Twindex's gain comes from serving 1.62\ensuremath{\times} as many requests.

\emph{Where each model's gain comes from.} One step-time model, $E(C) = F + c\cdot C$, explains the gains (Table~\ref{tab:18} in Appendix~\ref{app:extra}). Compression shrinks the index per request per GPU from 99.3 to 30.7 MiB on Qwen, from 43.2 to 21.6 MiB on V4.1 and from 336 to 79.1 MiB on GLM, which raises C. The cheaper scan lowers each request's step cost c, net of rescoring and fetches, from 53.9 to 41.3 \textmu{}s on Qwen and from 164.9 to 94.5 \textmu{}s on GLM. On V4.1, whose native index is already FP4, the gain comes from the larger batch.

\begin{table}[tp]
\centering
\footnotesize
\setlength{\tabcolsep}{3pt}
\caption{TPOT at equal concurrency and Twindex's added prefill time. TPOT is the mean decode step, HiSparse-Opt / Twindex. The second C is the highest concurrency that HiSparse-Opt fits in HBM on each model. Prefill is measured with one request at a time.}\label{tab:lat}
\begin{tabular}{@{}lrcrcc@{}}
\toprule
Model & C & TPOT (ms) & C & TPOT (ms) & Prefill, 1M / 16K \\
\midrule
Qwen & 128 & 30.5 / 29.8 & 861 & 69.9 / 61.3 & $<$0.1\% / $+$0.3\% \\
V4.1 & 128 & 22.6 / 24.1 & 630 & 43.9 / 45.8 & $+$0.2\% / $+$0.5\% \\
GLM & 128 & 61.1 / 33.8 & 232 & 79.8 / 47.4 & $+$0.4\% / $-$24\% \\
\bottomrule
\end{tabular}
\end{table}

\emph{TPOT at maximum throughput.} At maximum throughput, Twindex runs 1.78\ensuremath{\times} as many requests per step, and Qwen's decode step grows from 69.8 to 86.4 ms. The main-KV swap-in grows with C in both systems and lengthens Twindex's step by about 5 ms, and per-request work in the other layers adds the rest. On GLM, the swap-in dominates the growth, and on V4.1 it stays near 1 ms.

\emph{Throughput under a TPOT limit.} Under a median-TPOT limit, R is 1.11 at 40 ms and 1.44 at 100 ms (Figure~\ref{fig:6}, right), because a tight limit forces both systems to run small batches and a looser one lets Twindex fill its larger batch.

\emph{Time to first token.} Compressing the index adds at most 0.5\% to prefill on every model (Table~\ref{tab:lat}), and restoring only compressed keys cuts GLM's 16K-tail TTFT by 24\%. With prefill/decode disaggregation, Twindex's median TTFT, including admission to its larger decode batch, is 6.2 s against 5.9 s on V4.1 and 11.2 s against 11.0 s on Qwen.

\setlength{\parskip}{6pt}
\subsection{Model Quality}\label{sec:5.4}

Task scores compare the two systems as served, including HiSparse-Opt's top-k kernel (\S\ref{sec:5.2}), with greedy decoding unless noted, and selection metrics use the exact native top-k. Each item's score is its mean over 1--6 runs per system, and the 95\% interval resamples items (AIME25 problems with all their samples), paired across the two systems. AIME25 allows 32K thinking tokens, and LongGenBench writes 16K tokens after a 1M-token prompt. On V4.1 and GLM, the long-context benchmarks use only items of at least 500K tokens, because shorter items would have run the same native path in both systems. GLM runs fewer RULER, LongBench v2 and BABILong items (Table~\ref{tab:10}).

\emph{Long-context benchmarks.} Twindex's scores stay close to the baseline's across task types and lengths (Table~\ref{tab:10}): retrieval (RULER \citep{ref49}), long-document question answering (LongBench v2), reasoning over facts spread through a long text (BABILong \citep{ref53}), math with 32K-token thinking (AIME25 \citep{ref54}) and long-output generation (LongGenBench \citep{ref56}). On Qwen, Twindex's mean score is at most 2.5\% below the baseline's on each of the five benchmarks, relative to the baseline's score. The 95\% interval of the paired difference contains zero on four of the five. On AIME25, Twindex scores .010 higher, less than the .015 by which a second baseline run (.867) beats the first (.852). On V4.1 and GLM, Twindex's mean score is at most 3.1\% below the baseline's in our completed evaluations, and both systems score the same on LongBench v2.

\begin{table*}[tp]
\centering
\footnotesize
\caption{Benchmark quality of HiSparse-Opt (the baseline) and Twindex. Samples counts the scored items, times the samples per item for AIME25 and the runs per item for V4.1's LongGenBench. $\Delta$ is Twindex minus the baseline on Qwen, with its paired 95\% interval. V4.1 and GLM use subsets of the same items (\S\ref{sec:5.4}). GLM's AIME25 omits the two problems shorter than one 64-token index page, and both systems run the same 28.}\label{tab:10}
\setlength{\tabcolsep}{1.9pt}
\resizebox{\textwidth}{!}{%
\begin{tabular}{@{}llrcccc rc rc@{}}
\toprule
 & \multicolumn{6}{c}{Qwen} & \multicolumn{2}{c}{V4.1} & \multicolumn{2}{c}{GLM} \\
\cmidrule(lr){2-7}\cmidrule(lr){8-9}\cmidrule(lr){10-11}
Benchmark & Context & Samples & Baseline & Twindex & $\Delta$ & 95\% CI & Samples & Baseline / Twindex & Samples & Baseline / Twindex \\
\midrule
RULER & 1M & 384 & .875 & .886 & $+$.011 & [$-$.005, $+$.028] & 93 & .886 / .902 & 48 & .785 / .788 \\
LongBench v2 & 131K--1M & 203 & .400 & .392 & $-$.009 & [$-$.030, $+$.010] & 46 & .620 / .620 & 23 & .522 / .522 \\
BABILong & 1M & 150 & .515 & .527 & $+$.012 & [$-$.011, $+$.035] & 150 & .467 / .483 & 50 & .600 / .620 \\
AIME25 & $\le$33K & 30\,$\times$\,40 & .852 & .862 & $+$.010 & [$+$.001, $+$.020] & 30\,$\times$\,8 & .925 / .950 & 28\,$\times$\,8 & .723 / .701 \\
LongGenBench & 1M & 12 & .912 & .898 & $-$.014 & [$-$.039, $+$.000] & 12\,$\times$\,4 & .691 / .670 & 12 & .871 / .867 \\
\bottomrule
\end{tabular}}
\end{table*}

\emph{Exact top-k control.} HiSparse-Opt's top-k kernel misses some native top-512 entries on Qwen at long contexts (\S\ref{sec:5.2}), so we also run HiSparse-Opt with an exact top-512 on the same Qwen items. On all five benchmarks, the 95\% intervals of exact minus as-served and of Twindex minus exact contain zero (Table~\ref{tab:xtk}). On \GPU{}, an exact top-k from torch.topk raises HiSparse-Opt's mean TPOT from 69.9 to 88.0 ms at C = 861 and from 30.5 to 33.2 ms at C = 128. V4.1's top-k kernel returns the exact top-512 at 512K and 1M tokens, and GLM's top-k uses an exact sort.

\begin{table}[t]
\centering
\footnotesize
\setlength{\tabcolsep}{2.5pt}
\caption{Qwen scores with HiSparse-Opt's top-k as served (Base) and exact (Exact), and with Twindex, on the Qwen items of Table~\ref{tab:10}. The last two columns give paired 95\% intervals.}\label{tab:xtk}
\resizebox{\columnwidth}{!}{%
\begin{tabular}{@{}lccccc@{}}
\toprule
Benchmark & Base & Exact & Twindex & \begin{tabular}[b]{@{}c@{}}Exact\\$-$\,Base\end{tabular} & \begin{tabular}[b]{@{}c@{}}Twindex\\$-$\,Exact\end{tabular} \\
\midrule
RULER & .875 & .883 & .886 & [$-$.013,\,$+$.029] & [$-$.014,\,$+$.021] \\
LongBench v2 & .400 & .384 & .392 & [$-$.046,\,$+$.015] & [$-$.025,\,$+$.039] \\
BABILong & .515 & .513 & .527 & [$-$.039,\,$+$.037] & [$-$.024,\,$+$.051] \\
AIME25 & .852 & .864 & .862 & [$-$.001,\,$+$.028] & [$-$.017,\,$+$.011] \\
LongGenBench & .912 & .890 & .898 & [$-$.067,\,$+$.002] & [$-$.037,\,$+$.065] \\
\bottomrule
\end{tabular}}
\end{table}

\emph{Long generations.} Twindex's final set holds .943--.950 of the exact top-512 along 32,768-token outputs on Qwen and .957--.982 on V4.1 (Appendix~\ref{app:extra}.1). On Qwen it scores within .014 of the baseline on AIME25 and LongGenBench.

\emph{Selection fidelity.} Twindex's final set carries 99.87\% as much attention mass as the native top-512 over 128K--992K and 99.84\% at 992K, against 98.78\% for 2-bit quantization with the same 36 B and rescoring (Figure~\ref{fig:8}b in Appendix~\ref{app:B}). The attention mass of a set is the true softmax weight of its entries over the full context, averaged over heads and summed over 1,860 (layer, step) samples. The entries that Twindex misses rank low in the native order.

\subsection{Extending to Longer Contexts}\label{sec:long}

Twindex's gain keeps growing at 1.5M and 2M tokens, and its selection stays accurate there. We run this study on Qwen only. These lengths exceed Qwen's supported context, so we scale its RoPE \ensuremath{\times}8 with YaRN \citep{ref59} and use an index hot buffer of 4,096 rows per layer.

\emph{Throughput.} R reaches 1.562 at 1.5M with 1,304 requests against HiSparse-Opt's 622. At 2M, R is 1.684, with Twindex capped at 1,024 requests by the attention kernels' 32-bit token index, against HiSparse-Opt's 487.

\emph{Quality.} On RULER, Twindex's accuracy is within .001 of the baseline's at 1.5M and 2M, and both 95\% intervals, [$-$.030, $+$.028] and [$-$.022, $+$.021], contain zero. Its final set keeps 99.87\% and 99.79\% of the native top-512's attention mass.

\subsection{Ablation}\label{sec:5.6}

\looseness=-1 \emph{Shared component and residual compression.} Without the shared component, recall falls on all three models (\S\ref{sec:2.4}, Table~\ref{tab:abl}), and GLM's throughput falls by 15\%. A wider compressed key without it (54 B on Qwen, 45, 45, 36 and 24 B on V4.1's four layers and 36 B on GLM) recovers most of the recall (.939 on Qwen) but fits fewer requests, and throughput falls by 32\% on Qwen, 7\% on V4.1 and 9\% on GLM. An FP4 copy of each native row, rescored the same way, keeps nearly all of the native top-k but takes 68 B per entry on Qwen and 72 B on GLM, two and three times Twindex's 36 B and 24 B, and throughput falls by 25\% and 57\%. On V4.1, whose index is already FP4, this variant is HiSparse-Opt, with 17\% lower throughput.

\emph{Rescoring and index hot buffer.} Without rescoring, the compressed keys' own top-k holds only .699 of the native top-k on Qwen, .709 on V4.1 and .679 on GLM (Table~\ref{tab:abl}), with 1--2\% more throughput on Qwen and V4.1 and 12\% more on GLM. Without the index hot buffer, every candidate row crosses the host link, and throughput falls by 19\% on Qwen, 4\% on V4.1 and 45\% on GLM.

\begin{table}[tp]
\centering
\footnotesize
\caption{Effects of Twindex's components on recall and throughput. Each row removes one component or replaces it with a wider key (\S\ref{sec:5.6}). Recall is the share of the native top-k entries in the final selection on held-out documents. Thr.\ is the variant's decode throughput divided by Twindex's on the same node in the same session. Qwen's wider-key and FP4 rows cannot use the fused kernels, so they are divided by Twindex without fused kernels.}\label{tab:abl}
\setlength{\tabcolsep}{1.3pt}
\begin{tabular}{@{}lcccccc@{}}
\toprule
 & \multicolumn{2}{c}{Qwen} & \multicolumn{2}{c}{V4.1} & \multicolumn{2}{c}{GLM} \\
\cmidrule(lr){2-3}\cmidrule(lr){4-5}\cmidrule(lr){6-7}
Variant & Recall & Thr. & Recall & Thr. & Recall & Thr. \\
\midrule
Twindex & .949 & 1.00 & .994 & 1.00 & .997 & 1.00 \\
$-$ shared component & .831 & 1.02 & .899 & 1.00 & .710 & 0.85 \\
$-$ shared, wider key & .939 & 0.68 & .936 & 0.93 & .848 & 0.91 \\
$-$ residual (FP4) & .998 & 0.75 & 1.000 & 0.83 & .983 & 0.43 \\
$-$ rescoring & .699 & 1.01 & .709 & 1.02 & .679 & 1.12 \\
$-$ hot buffer & .949 & 0.81 & .994 & 0.96 & .997 & 0.55 \\
\bottomrule
\end{tabular}
\end{table}

\subsection{Sensitivity Analysis}\label{sec:sens}

The sensitivity studies quantify the tradeoffs around the deployed D, M and $\alpha$ (Tables~\ref{tab:alpha} and~\ref{tab:sens} and Figure~\ref{fig:14} in Appendix~\ref{app:extra}). On Qwen, a 54 B compressed key (D = 96) with the shared component finds .991 of the native top-512 against .949 at 36 B (Table~\ref{tab:B.2} in Appendix~\ref{app:B}). The 54 B key without the shared component has 32\% lower throughput than Twindex in the same session (Table~\ref{tab:abl}). A 64-dimension key on every GLM layer has 11\% lower throughput than Twindex's in the same session. More candidates cost time, because Twindex reads and rescores one native row for each. On Qwen, the final set keeps 99.60\% of the native top-512's attention mass at M = 768 and 99.95\% at 2,048, and in one paired session R is 1.49 at M = 768, 1.46 at the deployed 1,280 and 1.42 at 1,792. A smaller D needs more candidates, and on V4.1 recall .964 takes 922 candidates at D = 80 and 4,438 at D = 48. Blending the query metric with the identity raises Qwen's recall by .053 or more over $\alpha = 0$ and $\alpha = 1$, and GLM does best with $\alpha = 0$.

\section{Related Work}\label{sec:6}

\textbf{Indexer cache.} Table~\ref{tab:rw} compares the designs. HiSparse \citep{ref15}, ECHO \citep{ref16}, ESS \citep{ref17}, SAC \citep{ref18} and HiSparse v2 \citep{ref75} keep the native index in HBM. IndexCache \citep{ref9} reuses the previous indexer layer's top-k, and LatentIndex \citep{ref77} shares one latent index across layers, evaluated at up to 128K tokens. Self-Indexing \citep{ref70} replaces the index keys with 1-bit signs, and FlashMemory-DeepSeek-V4 \citep{ref10} trains a new indexer that fetches candidates for the native one. We re-implement SAKI and Self-Indexing from their papers (Appendix~\ref{app:B}.9). At the same bytes and M, SAKI's fit \citep{ref71} finds .874 of Qwen's native top-512 against .949 for Twindex, and Self-Indexing's 1-bit signs find .37--.51. Served on V4.1, Self-Indexing reaches R~=~0.96, with RULER accuracy .695 against Twindex's .902 on the same 93 items.

\textbf{Sparse attention, KV offloading and serving.} Quest \citep{ref1}, NSA \citep{ref2}, MoBA \citep{ref3} and DSA \citep{ref4} keep every past token selectable, unlike linear attention \citep{ref11,ref12} and sliding windows \citep{ref13,ref14}. HISA \citep{ref8} filters blocks before native-indexer refinement, and MISA \citep{ref69} routes indexer heads, with a variant that re-ranks an enlarged candidate set with the full native indexer. InfiniGen \citep{ref19}, ShadowKV \citep{ref20}, RetrievalAttention \citep{ref21}, MagicPIG \citep{ref22}, ParisKV \citep{ref23} and RetroInfer \citep{ref24} offload KV state to host memory and identify relevant entries through prediction, search or sampling. vLLM \citep{ref40}, SGLang \citep{ref41}, DistServe \citep{ref42}, Splitwise \citep{ref43} and Mooncake \citep{ref44} manage, share and move caches across requests and instances.

\textbf{Key compression and re-ranking.} KIVI \citep{ref27}, KVQuant \citep{ref28} and Palu \citep{ref29} reconstruct keys and values. Loki \citep{ref30}, RaBitQCache \citep{ref31}, SAKI, SparQ \citep{ref32} and PQCache \citep{ref33} select entries from approximate scores, and SAKI also subtracts a calibration mean of the keys and fits the compression in the query metric. Twindex applies both to the indexer cache and adds native rescoring and an index hot buffer. Vector search \citep{ref34,ref35,ref36,ref37,ref38,ref39} screens with compressed vectors and re-ranks a short list with finer vectors or exact distances, as Twindex does with native rows.

\section{Conclusion}\label{sec:7}

Twindex compresses the indexer cache of sparse-attention models at million-token contexts. At 1M tokens, Twindex reaches 1.44\ensuremath{\times}, 1.25\ensuremath{\times} and 2.54\ensuremath{\times} HiSparse-Opt's decode throughput on Qwen, V4.1 and GLM, and mean benchmark scores at most 3.1\% lower in our completed evaluations. Next, we will measure Twindex with a cache tier below host memory (\S\ref{sec:5.3}).

\clearpage
\label{mainend}

\bibliographystyle{plainnat}
\bibliography{main}

\clearpage
\beginappendix

\appendix
\label{appstart}
\counterwithin{figure}{section}
\counterwithin{table}{section}
\renewcommand{\theHfigure}{app.\thefigure}\renewcommand{\theHtable}{app.\thetable}

\section{Additional baseline optimizations}\label{app:A}

Table~\ref{tab:A.1} lists the smaller optimizations behind HiSparse-Opt (\S\ref{sec:5.2}), beyond the two fixes that \S\ref{sec:5.2} describes. Each optimization applies to both systems wherever both systems have the component. When a Qwen host arena exceeds the driver's limit of about 504 GiB for one pinned registration, both systems register it in pieces of at most 256 GiB. On V4.1, both systems store the host arena once per node instead of once per GPU, which makes it 4\ensuremath{\times} smaller, and freeze surviving Python objects every 10 s, which removes garbage-collection stalls of 135--175 ms. Together they shorten HiSparse-Opt's step at C = 640 from 43.75 to 43.53 ms. We measure the Effect column on HiSparse-Opt on \GPU{} at 1M tokens.

\begin{table*}[tp]
\centering
\footnotesize
\caption{Smaller optimizations behind HiSparse-Opt. The Effect column is measured on HiSparse-Opt (\GPU{}, 1M tokens).}\label{tab:A.1}
\begin{tabular}{@{}>{\raggedright\arraybackslash}p{\dimexpr 0.306\linewidth-2\tabcolsep\relax}>{\raggedright\arraybackslash}p{\dimexpr 0.306\linewidth-2\tabcolsep\relax}>{\raggedright\arraybackslash}p{\dimexpr 0.307\linewidth-2\tabcolsep\relax}>{\raggedright\arraybackslash}p{\dimexpr 0.081\linewidth-2\tabcolsep\relax}@{}}
\toprule
Problem found & Fix & Effect & Model \\
\midrule
Decode CUDA graphs cover only batches up to 512, but full-HBM concurrency is 861--1,532. Eager decode spends 131--155 ms of host time per step & Capture the whole decode step as one graph at batch = C & Host time 0.9--1.1 ms per step, step 159.6 \ensuremath{\rightarrow} 138.1 ms & Qwen \\
KV swap-in serialized with compute & Two-group schedule: one half-batch swaps in on a side stream while the other half computes, all in one graph & Step 111.3 \ensuremath{\rightarrow} 98.1 ms (\ensuremath{-}12\%) & Qwen \\
Sparse decode attention runs as three kernels (compaction, sequence-offset setup, variable-length attention) & One fused kernel, bitwise equal & About \ensuremath{-}4.5 ms per step & Qwen \\
Greedy sampling all-gathers the full logits across ranks & Per-rank argmax first, then gather {[}C, 2{]} & \ensuremath{-}1.12 ms per step, graph pool \ensuremath{-}696 MiB & Qwen \\
The position-to-slot map is sized for the 1M maximum (4.0 MiB per request per GPU). Each step reads only about 2K of its entries & Compact map of {[}request, 2,052{]} & 4.0 MiB per request returned to concurrency & Qwen \\
The launcher's vision warmup leaves a hidden 710 MiB CUDA context on GPU 0. GPU 0's peak sets the largest C that meets the HBM limit & Keep the parent process off the GPU & Rank-0 surplus +678 \ensuremath{\rightarrow} \ensuremath{-}64 MiB & Qwen \\
A stream-ordering call pins graph memory. The dense capture reserves full width & Remove that call, add a private pool, clamp the capture width & Graph pool 1,282 \ensuremath{\rightarrow} 320 MiB per GPU, concurrency 216 \ensuremath{\rightarrow} 227, Q +2.6\% & V4.1 \\
Unfused swap-in exposes 7.3 ms per step & Fused swap kernel & \ensuremath{-}5 ms per step & V4.1 \\
Address translation bounds the host gathers of small working-set entries & Host working-set arena in managed memory & Q +5.8\% & V4.1 \\
Admission stalls above about 400 streams. The token-streaming path tops out near 6K chunks/s & Stream output every 16 tokens, incrementally & Concurrency above 400 becomes reachable & V4.1 \\
\bottomrule
\end{tabular}
\end{table*}

\textbf{Top-k correctness.} SGLang's original top-k kernel for Qwen puts the candidates of the score bin that holds the top-k threshold into a 4,096-entry buffer and silently drops the rest. At 512K tokens and beyond, some calls therefore return a top-512 that misses a higher-scoring row. How often depends on the document, the length and the layer: 6.7\% of the checked Qwen decode calls on identical scores on \GPU{}, and up to 12.4\% at 992K on one trace. HiSparse-Opt keeps the original kernel, so its selection can differ from the exact top-k (\S\ref{sec:5.2}). At 992K tokens, it returns on average 4.3 entries per call (0.85\%) that differ from the exact top-512, and 93\% of them have exact ranks 513--640. On Qwen, Twindex's final top-k ranks at most 1,408 candidates: $M = 1{,}280$ plus at most 128 blocks of $\mathcal{A}$ (the first 64 and the most recent 64, \S\ref{sec:3.3}). It never overflows the buffer, so its final top-k is exact over its candidates. V4.1's top-k kernel holds at most 851 entries of the threshold bin at 1M tokens and 506 at 512K, within its 2,048-entry limit, so it returns the exact top-512 at both lengths. GLM's top-k uses an exact sort.

\section{Index-key structure: derivations and full statistics}\label{app:B}

This appendix gives the derivations and full statistics behind \S\ref{sec:2.4} and \S\ref{sec:3.2}.

\textbf{Setup.} Qwen3.8-Flash-Next, 12 indexer layers, 4-token block keys. RoPE rotates dimensions 0--63 at the deployment's YaRN frequencies (\ensuremath{\times}4 at 1M context, \ensuremath{\times}8 at 1.5--2M). Dimensions 64--127 are not rotated. The data splits are:

\begin{itemize}
\item Calibration set: five captures of four documents at 512K and 1M tokens, used for fitting.
\item Development set: 19 captures of 18 documents at 512K and 992K. We used them to choose D, M and $\alpha$, never for fitting.
\item Held-out set: 31 captures in 30 document clusters after removing shared prefixes, from 10 domains at 128K--992K tokens, with 1,860 (decode step, layer) samples: 5 decode steps \ensuremath{\times} 12 indexer layers per capture. We never used the held-out set for fitting or for choosing D, M and $\alpha$.
\item Long-context set: 11 captures at 1.5M--2M from 6 of the 10 domains, with 660 (decode step, layer) samples.
\end{itemize}

All confidence intervals are 95\% document-cluster bootstrap intervals.

\textbf{B.1 Shared component.} Write each index key as $k_t = R(t\theta)(\mu + z_t)$, where $R(t\theta)$ is the RoPE rotation at position $t$, $\mu$ is the shared pre-rotation mean and $z_t$ is a deviation with $\mathrm{E}[z] = 0$. Then $\mathrm{E}[k_t] = R(t\theta)\,\mu$. For an estimate $\mu_c$ of the shared mean,
\[
\mathrm{E}\lVert k - R(t\theta)\,\mu_c\rVert^2 = \lVert \mu_d - \mu_c\rVert^2 + \operatorname{tr}\Sigma_d,
\]
where $\mu_d$ and $\Sigma_d$ are the document's pre-rotation mean and covariance. We call the mean squared key norm $\mathrm{E}\lVert k\rVert^2$ the key energy. The fraction of key energy that $\mu_c$ explains is therefore $(2\mu_c\cdot\mu_d - \lVert\mu_c\rVert^2)/\mathrm{E}\lVert k\rVert^2$. On the held-out set, the per-document means align with $\mu_c$ (cosine .905 {[}.880, .927{]}).

\textbf{Why a shared component fitted offline suffices.} The shared component does not need to be the centre of each document's keys. Twindex computes the contribution $q\cdot R(t\theta)\,\mu_c$ of the shared component exactly from the position, and compresses only the residual. The residual energy of a document $d$ with $n_d$ keys is the sum of its squared residual norms. Because $R(t\theta)$ preserves length, this energy splits exactly into two parts, $\sum_t \lVert k_t - R(t\theta)\,\mu_c\rVert^2 = \sum_t \lVert u_t - \mu_d\rVert^2 + n_d\lVert \mu_d - \mu_c\rVert^2$, where $u_t$ is the key before rotation. This split is the summed form of the identity above. The first part does not depend on $\mu_c$, so any fixed $\mu_c$ pays only the second part. Across documents, the expected second part $\mathrm{E}_d\lVert \mu_d - \mu_c\rVert^2$ is smallest when $\mu_c$ is the average of the document means. The calibration fit estimates this average.

\begin{figure}[t]
\centering
\includegraphics[width=\columnwidth]{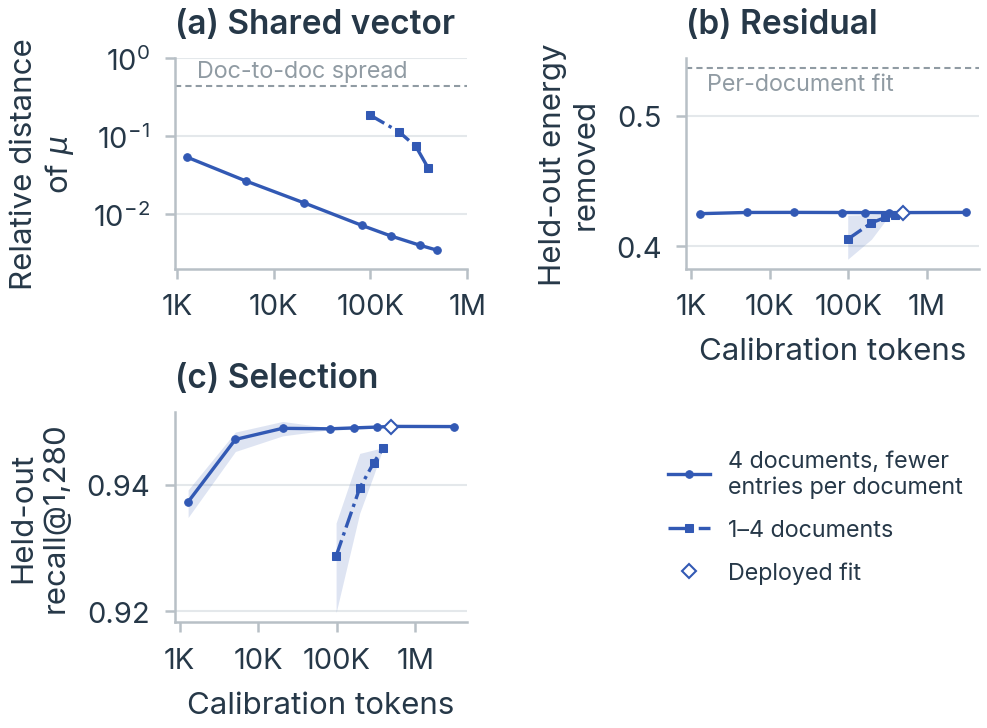}
\caption{Calibration size of the shared-component fit, evaluated on the held-out set: (a) distance from the deployed fit, (b) held-out key energy removed, (c) recall@1280. More calibration documents help. More tokens per document do not.}\label{fig:B.1}
\end{figure}

\textbf{Calibration size.} The fit needs few tokens per document but several documents (Figure~\ref{fig:B.1}). Two independent samples of the deployed size give fits of $\mu_c$ that differ by 0.35\% of its length, more than 100 times less than the spread of the held-out documents' own means, so more tokens per document leave the held-out residual unchanged: with 1,024 entries per capture (20K tokens in all), held-out recall is within 0.001 of the deployed fit. More documents do help: a fit on a single calibration document has 0.020 lower recall (95\% CI 0.017 to 0.024).

The fixed $\mu_c$ keeps most of a per-document fit's gain: it removes 79\% of the key energy that a per-document fit removes (.426 of .537, Table~\ref{tab:B.1}).

\begin{table*}[tp]
\centering
\footnotesize
\caption{Share of held-out key energy and variance that each predictor explains (Qwen, held-out set, 95\% intervals in brackets).}\label{tab:B.1}
\begin{tabular}{@{}>{\raggedright\arraybackslash}p{\dimexpr 0.296\linewidth-2\tabcolsep\relax}>{\raggedright\arraybackslash}p{\dimexpr 0.250\linewidth-2\tabcolsep\relax}>{\raggedright\arraybackslash}p{\dimexpr 0.263\linewidth-2\tabcolsep\relax}>{\raggedright\arraybackslash}p{\dimexpr 0.191\linewidth-2\tabcolsep\relax}@{}}
\toprule
Predictor (held-out set) & Energy explained, all dims & Centred variance, RoPE dims & Centred $q\cdot k$ energy \\
\midrule
Constant mean & .209 {[}.197, .220{]} & .000 & .000 \\
Shared component (ours) & .426 {[}.408, .444{]} & .485 {[}.446, .527{]} & .262 {[}.219, .310{]} \\
Shuffled positions (control) & \ensuremath{-}.040 {[}\ensuremath{-}.055, \ensuremath{-}.027{]} & \ensuremath{-}.678 & \ensuremath{-}.330 \\
Per-document fit (upper reference) & .537 {[}.491, .588{]} & .582 {[}.534, .637{]} & .305 {[}.250, .369{]} \\
\bottomrule
\end{tabular}
\end{table*}

\begin{figure}[t]
\centering
\includegraphics[width=\columnwidth]{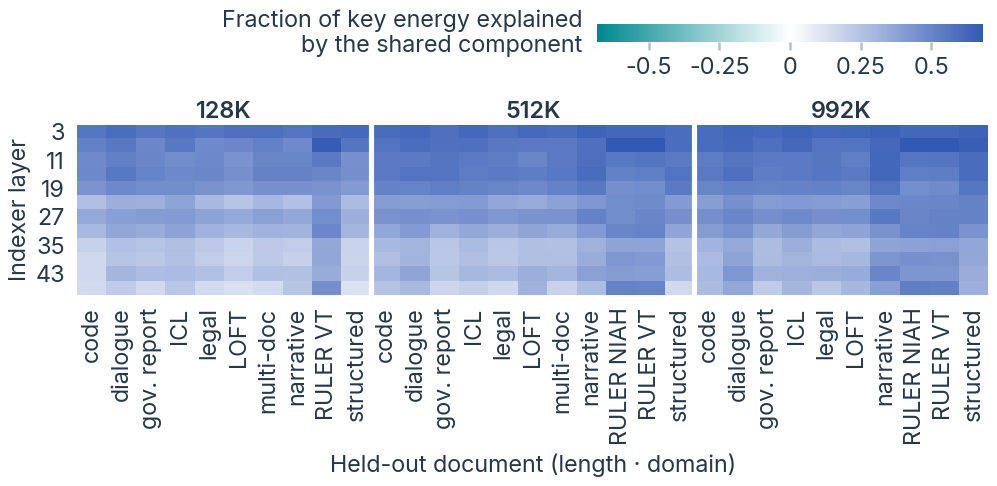}
\caption{The shared component helps in every document and every indexer layer: the key energy it explains is positive in every cell. Qwen, held-out documents by context length and domain.}\label{fig:B.2}
\end{figure}

The gain of the shared component should not depend on one length or one layer. The key energy that it explains grows with context length: .372 (128K), .436 (512K), .468 (992K), .505 (1.5M) and .505 (2M). The explained energy is positive at every layer, from .61 at layer 3 to .27 at layer 47 (Figure~\ref{fig:B.2}).

\textbf{B.2 Optimal residual compression.} Let $S_r$ be the covariance of the residuals and $\Sigma_q = \mathrm{E}[qq^\top]$ the second moment of the queries, taken about zero. The fit centres the residuals before the eigendecomposition. For $q$ independent of $r$, a centred residual $r$ and any rank-$D$ linear compression $\hat r = Pr$,
\[
\mathrm{E}\big(q\cdot(r - Pr)\big)^2 \ \ge\ \sum_{i>D} \lambda_i\big(\Sigma_q^{1/2} S_r\, \Sigma_q^{1/2}\big),
\]
with equality at $P = \Sigma_q^{-1/2} U_D U_D^\top \Sigma_q^{1/2}$, where $U_D$ holds the top-$D$ eigenvectors of $\Sigma_q^{1/2} S_r \Sigma_q^{1/2}$. Twindex's compression is this optimum for its fitted metric $G$ before quantization. On all 12 layers, the compression's error equals the bound. Averaged over positions whose rotation phases cover the circle uniformly, a rotating mean adds $(\lvert\mu_i\rvert^2/2)\,I_2$ to the key second moment in each fast-rotating plane. Removing the rotating mean frees these dimensions for content. SAKI \citep{ref71} derives the same optimum for attention keys centred before rotation (its Proposition 1).

\begin{table*}[tp]
\centering
\footnotesize
\caption{Recall of the native top-512 within 1,280 candidates by compressed width D (Qwen, held-out set, NVFP4 compressed keys of 9D/16 bytes per entry).}\label{tab:B.2}
\begin{tabular}{@{}>{\raggedright\arraybackslash}p{\dimexpr 0.302\linewidth-2\tabcolsep\relax}>{\raggedright\arraybackslash}p{\dimexpr 0.100\linewidth-2\tabcolsep\relax}>{\raggedright\arraybackslash}p{\dimexpr 0.100\linewidth-2\tabcolsep\relax}>{\raggedright\arraybackslash}p{\dimexpr 0.100\linewidth-2\tabcolsep\relax}>{\raggedright\arraybackslash}p{\dimexpr 0.101\linewidth-2\tabcolsep\relax}>{\raggedright\arraybackslash}p{\dimexpr 0.100\linewidth-2\tabcolsep\relax}>{\raggedright\arraybackslash}p{\dimexpr 0.100\linewidth-2\tabcolsep\relax}>{\raggedright\arraybackslash}p{\dimexpr 0.100\linewidth-2\tabcolsep\relax}@{}}
\toprule
D (bytes per entry) & 16 (9 B) & 32 (18 B) & 48 (27 B) & 64 (36 B) & 80 (45 B) & 96 (54 B) & 128 (72 B) \\
\midrule
Recall@1280, with shared component & .660 & .817 & .902 & .949 & .977 & .991 & .9996 \\
Recall@1280, without shared component & .520 & .656 & .752 & .831 & .890 & .939 & .9925 \\
Recall@1280, constant mean & .545 & .686 & .785 & .863 & .920 & .965 & .9989 \\
\bottomrule
\end{tabular}
\end{table*}

\begin{figure}[t]
\centering
\includegraphics[width=\columnwidth]{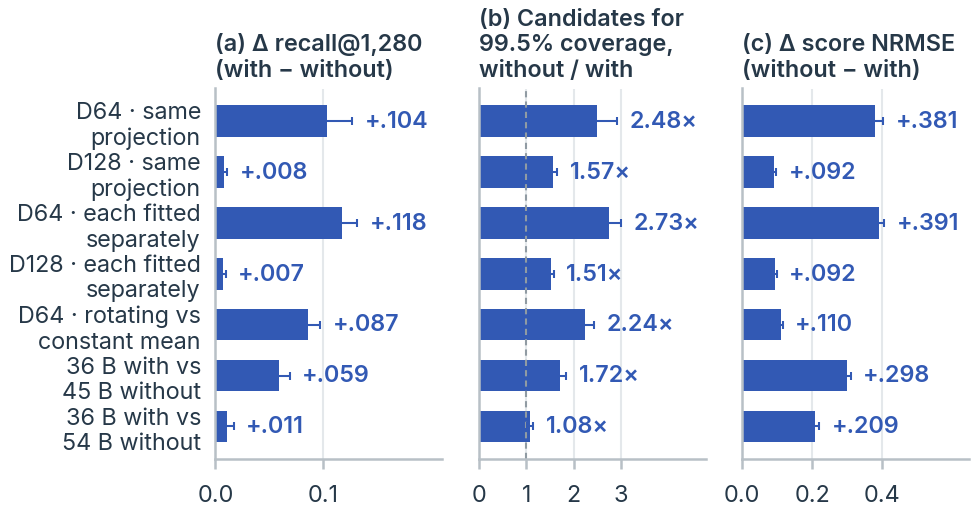}
\caption{Causal ablation of the shared component (held-out documents, M = 1,280): each row compares compressed keys with and without it. (a) Recall gain. (b) Ratio of candidates needed for 99.5\% coverage. (c) Reduction in the normalized RMS error of the scores.}\label{fig:B.3}
\end{figure}

\textbf{B.3 Causal ablation (held-out set, D = 64, M = 1,280, Figure~\ref{fig:B.3}).} These ablations test whether the shared component itself causes the gain, rather than another fitting choice.

\begin{figure}[t]
\centering
\includegraphics[width=\columnwidth]{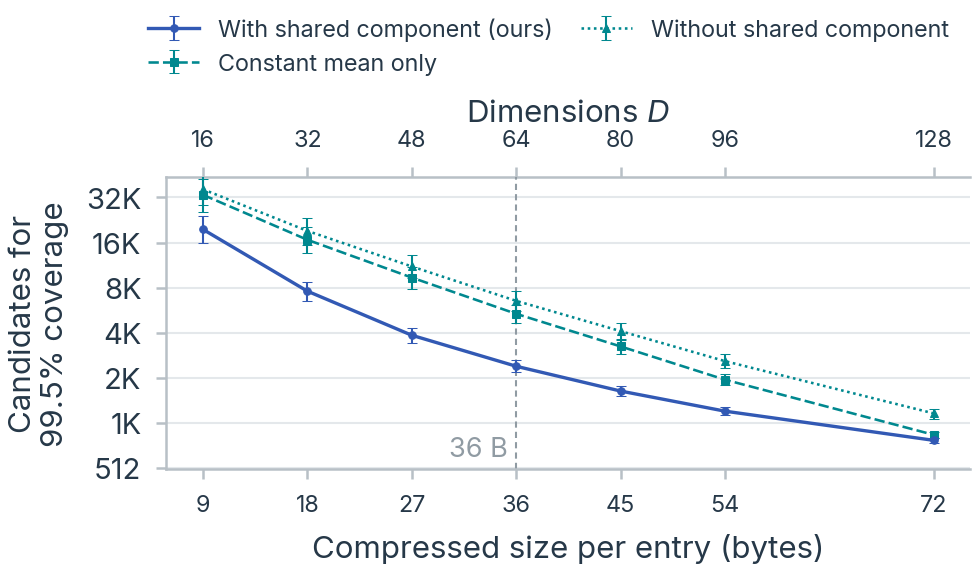}
\caption{With the shared component, the compressed key needs fewer candidates at every size (Qwen, held-out documents). The plot shows the candidates needed for 99.5\% coverage of the native top-512 against compressed size. The dashed line marks the deployed 36 B compressed key.}\label{fig:B.4}
\end{figure}

\begin{itemize}
\item Shared component on vs. off with the same projection: +.104 {[}+.083, +.127{]} recall, and a 2.48\ensuremath{\times} smaller M for 99.5\% coverage of the native top-512. Figure~\ref{fig:B.4} shows the smaller M at every compressed size.
\item Rotating shared component vs. constant mean: +.087 recall and a 2.24\ensuremath{\times} smaller M.
\item 36 B with the shared component vs. 54 B without: +.011 {[}+.005, +.017{]} recall.
\item On the development set, which we used to choose D, M and $\alpha$, the shared component adds +.22 recall at D = 64. On the long-context set, it adds +.20.
\end{itemize}

\begin{figure}[tp]
\centering
\includegraphics[width=\columnwidth]{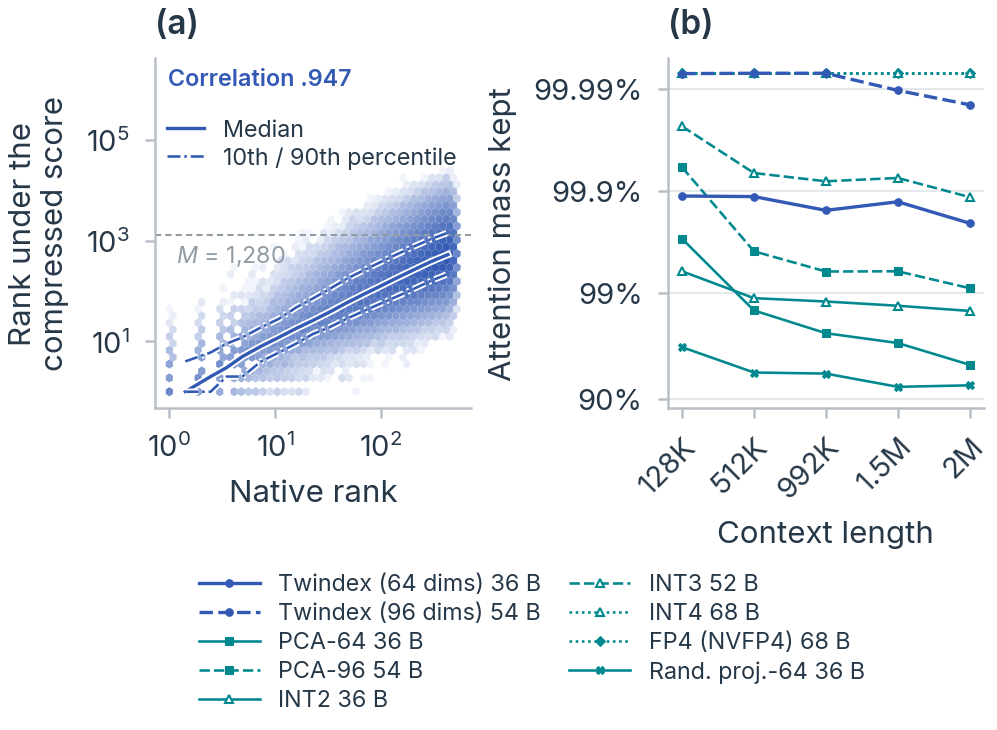}
\caption{Ranking fidelity and attention mass kept, by index format (Qwen). (a) Rank under the compressed score against native rank (31 held-out documents, 128K--992K). (b) Share of the native top-512's attention mass kept after rescoring 1,280 candidates, by index format (logit axis). The PCA arms are fitted on each document itself.}\label{fig:8}
\end{figure}

\textbf{B.4 Selection stability.} This check shows which native entries Twindex keeps after rescoring. Compressed scores correlate with native scores at .947. Twindex selects M = 1,280 candidates and rescores them at native precision. After rescoring, Twindex keeps .998 of the native top-512 entries at native ranks 1--128, .978 at ranks 129--256 and .910 at ranks 257--512 (overall .949 {[}.942, .956{]}). The differences therefore concentrate in the lower half of the top-512. This concentration is consistent with Observation 1 (\S\ref{sec:2.4}). Figure~\ref{fig:8} shows the ranks and the attention mass that each 36 B index format keeps.

\textbf{B.5 Held-out evaluation.} Except for the development-set result in B.3 and the explicitly labeled per-document fits, we measure every statistic in this appendix on documents that we never used for fitting or for choosing D, M and $\alpha$. These documents are 30 held-out clusters from 10 domains at 128K--992K and 11 long-context captures at 1.5M--2M. The gain from the shared component holds in every domain, layer and length (Figure~\ref{fig:B.2}). Averaged per document, no held-out document scores lower with the shared component.

\textbf{B.6 RoPE frequencies.} We fit and apply the shared component in the rotary frame that the model uses at inference, including YaRN scaling (for Qwen, \ensuremath{\times}4 at 1M and \ensuremath{\times}8 at 1.5--2M). Our implementation reads the model's RoPE configuration at load time. The shared component therefore always matches the deployed frequencies.

\textbf{B.7 Projection metric.} This comparison tests which part of the query metric produces the recall gain of \S\ref{sec:3.2}. We compare three ways of fitting the D = 64 residual compression: plain PCA (G = I), PCA in the query metric alone (\ensuremath{\alpha} = 0) and the blend that Twindex uses (\ensuremath{\alpha} = 0.5), each with and without averaging over position shifts, with the same calibration data, shared component and NVFP4 quantization. Table~\ref{tab:B.3} reports the recall of the native top-512 within 1,280 candidates, the candidates needed for 99.5\% recall and the normalized RMS error of the scores.

\begin{table*}[tp]
\centering
\footnotesize
\caption{Fits of the D = 64 residual compression: plain PCA, the query metric alone and the blend, each with and without shift averaging (Qwen, NVFP4).}\label{tab:B.3}
\begin{tabular}{@{}>{\raggedright\arraybackslash}p{\dimexpr 0.202\linewidth-2\tabcolsep\relax}>{\raggedright\arraybackslash}p{\dimexpr 0.123\linewidth-2\tabcolsep\relax}>{\raggedright\arraybackslash}p{\dimexpr 0.145\linewidth-2\tabcolsep\relax}>{\raggedright\arraybackslash}p{\dimexpr 0.133\linewidth-2\tabcolsep\relax}>{\raggedright\arraybackslash}p{\dimexpr 0.124\linewidth-2\tabcolsep\relax}>{\raggedright\arraybackslash}p{\dimexpr 0.145\linewidth-2\tabcolsep\relax}>{\raggedright\arraybackslash}p{\dimexpr 0.128\linewidth-2\tabcolsep\relax}@{}}
\toprule
Projection & Held-out recall & Held-out candidates & Held-out score error & 1.5M--2M recall & 1.5M--2M candidates & 1.5M--2M score error \\
\midrule
Plain PCA & .895 {[}.877, .911{]} & 4,388 & .463 & .834 {[}.814, .854{]} & 7,557 & .428 \\
Plain PCA + shift averaging & .896 {[}.879, .912{]} & 4,337 & .461 & .837 {[}.816, .857{]} & 7,334 & .425 \\
Query metric only (\ensuremath{\alpha} = 0) & .839 {[}.822, .857{]} & 6,737 & .497 & .774 {[}.747, .800{]} & 12,678 & .476 \\
Query metric only + shift averaging & .871 {[}.857, .886{]} & 5,258 & .443 & .823 {[}.800, .846{]} & 8,822 & .419 \\
Blend (\ensuremath{\alpha} = 0.5) & .950 {[}.943, .956{]} & 2,443 & .389 & .895 {[}.883, .910{]} & 4,337 & .385 \\
Blend + shift averaging (Twindex) & .949 {[}.942, .956{]} & 2,441 & .382 & .912 {[}.903, .924{]} & 3,710 & .366 \\
\bottomrule
\end{tabular}
\end{table*}

\begin{itemize}
\item Fitting in the blended query metric beats plain PCA at every length: the paired gain is +.054 {[}.042, .065{]} recall on the held-out set and +.076 {[}.062, .091{]} on the long-context set.
\item Blending with the identity makes the metric reliable. The calibration query second moment has rank 100 of 128 in every layer, so the query metric alone ($\alpha = 0$) gives no weight to directions that held-out queries do use, and its recall stays below that of plain PCA.
\item Averaging over position shifts matters beyond the calibrated lengths: it leaves held-out recall unchanged but adds +.017 {[}.014, .021{]} recall at 1.5M--2M tokens.
\end{itemize}

\textbf{B.8 Fitting the query metric offline.} Twindex fits the query metric once, from calibration queries, and serves every request with the same compression. Here $\Sigma_q = \mathrm{E}[qq^\top]$ is the query second moment of B.2. For a rank-$D$ compression $P$, $E_P = (I-P)\,S_r\,(I-P)^\top$ is the second moment of its error on the centred residuals. Two facts make one offline fit sound.

\begin{itemize}
\item The score error depends on the queries only through $\Sigma_q$. The compression changes the score of query $q$ on key residual $r$ by $q^\top(I-P)\,r$. Every query scores every key, so the mean squared error over any set of queries and keys is $\mathrm{tr}(\Sigma_q E_P)$. Individual queries do not enter the fit.
\item One fixed compression is best fitted to the mean query second moment when documents share one residual second moment $S_r$. Let $\Sigma_j$ be the query second moment of document $j$ and $\bar\Sigma$ its mean over documents. With a common $S_r$, the error of a fixed compression averaged over documents is $\mathrm{tr}(\bar\Sigma E_P)$, and by B.2 the compression fitted with $\bar\Sigma$ minimizes it. On document $j$, this compression exceeds the best compression for $\Sigma_j$ by a term $\delta_j \ge 0$, which is all that a per-request fit can recover. This term plays the role of the distance term of the shared component in B.1.
\end{itemize}

The blend with the identity, $G = (1-\alpha)\,\Sigma_q/s + \alpha I$, gives every direction a weight of at least $\alpha$. With 100 calibration queries in 128 dimensions, the sample $\Sigma_q$ is singular, and the query metric alone gives zero weight to directions that no calibration query uses. The weak directions matter on held-out documents. After shift averaging, 5.8 directions per layer have less than 1\% of the mean eigenvalue of the calibration $\Sigma_q$. They hold 0.03\% of the calibration query energy but 2.5\% of the held-out query energy.

\begin{figure}[tp]
\centering
\includegraphics[width=\columnwidth]{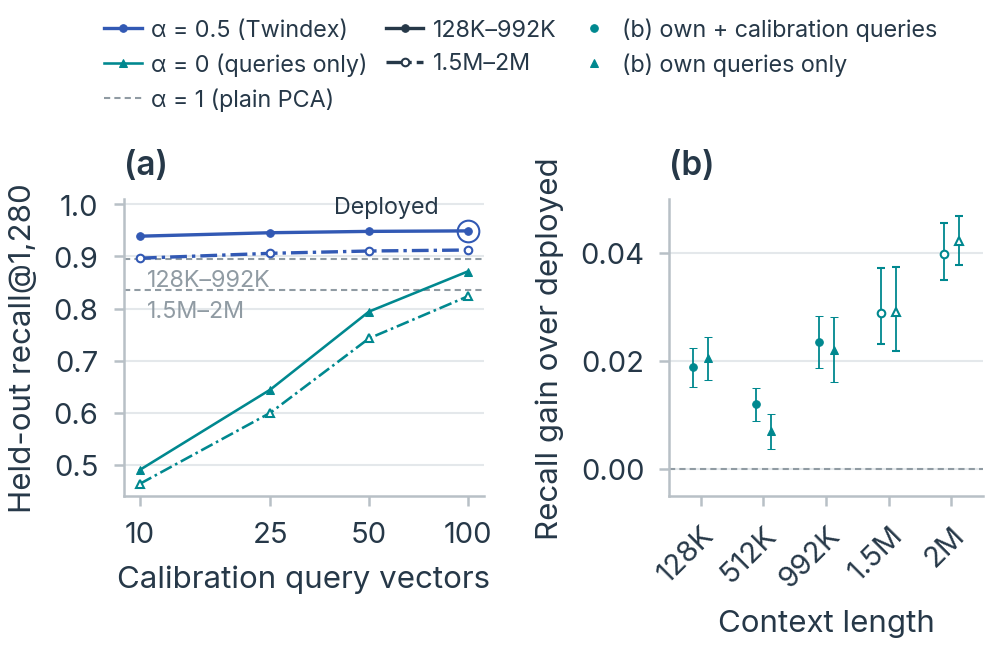}
\caption{A per-document refit of the query metric adds little recall over one offline fit (Qwen, D = 64). (a) Recall@1280 against the number of calibration query vectors, for the blend and the query metric alone. (b) Recall gain when the projection is refitted per document.}\label{fig:B.5}
\end{figure}

Fifty calibration queries reach the plateau (Figure~\ref{fig:B.5}a): with $\alpha = 0.5$, they come within .001 of 100 at 128K--992K. The query metric alone ($\alpha = 0$) needs many queries. With 100 it reaches .871, below plain PCA at .896.

A per-request fit gains little (Figure~\ref{fig:B.5}b). Refitting the projection on each held-out document's own decode queries from its other four decode steps, mixed with the calibration queries ($\alpha = 0.5$), raises recall by .018 {[}.015, .021{]} on average at 128K--992K and .035 {[}.030, .041{]} at 1.5M--2M.

V4.1 shows the same. Using 200 of the 640 calibration query vectors changes recall by $-$.001, and $\alpha = 0.25$ and 0.5 tie. A per-request fit adds little beyond a position effect that rotating the calibration metric to the served length recovers: .019 of .034 at 992K. GLM keeps $\alpha = 0$. It does best even with 100 of its 27,648 calibration query vectors (.976 against .960 for $\alpha = 0.5$ at 992K), and 1,600 of them come within .001 of all of them at half the served candidate budget, where recall is not saturated. A per-request fit raises recall at the served budget by .0001 at 512K and .002 at 992K, all of which rotating the calibration metric to the served length recovers. We therefore keep one offline fit per model.

\textbf{B.9 Comparison with SAKI and Self-Indexing.} We re-implement both methods from their papers, fit them on the same calibration data and score them on the same held-out documents as Twindex (Table~\ref{tab:B.4}). For SAKI \citep{ref71}, we use its latest form: a constant mean of the keys before rotation, the query metric alone averaged over 128 rotation offsets, and the centred key covariance. At Twindex's bytes, it uses Twindex's compressed width and NVFP4 quantization, so only the fit differs. For the rank-32 reference, we store the compressed keys in FP16 (64\,B per entry). For Self-Indexing \citep{ref70}, we store the 128 sign bits of a randomized Hadamard transform of each key and its norm, and score each index head by the sign agreement above 64, scaled by the norm and weighted by the model's head weights. Candidate recall is the share of the native top-k within the served candidates after native rescoring. Own recall is the share within each method's own top-k, without rescoring.

\begin{table*}[tp]
\centering
\footnotesize
\caption{Recall of the native top-k for Twindex, SAKI and Self-Indexing (held-out documents). Qwen: top-512 within M = 1,280, 128K--992K. V4.1: top-512 within the served M at layers 2, 8 and 14 (own recall also includes layer 20), in an NVFP4 replay (36--45 B). GLM: top-2,048 within the served $M_l$, 992K. 95\% intervals for Qwen: Twindex .949 [.942, .956], SAKI at Twindex's bytes .874 [.857, .891].}\label{tab:B.4}
\begin{tabular}{@{}lccccccc@{}}
\toprule
 & & \multicolumn{2}{c}{Qwen} & \multicolumn{2}{c}{V4.1} & \multicolumn{2}{c}{GLM} \\
\cmidrule(lr){3-4}\cmidrule(lr){5-6}\cmidrule(lr){7-8}
Method & Bytes per entry & Candidates & Own & Candidates & Own & Candidates & Own \\
\midrule
Twindex & 36 / 36--45 / 24 & .949 & .699 & .998 & .740 & .997 & .679 \\
SAKI, Twindex's bytes & same & .874 & .618 & .984 & .692 & .996 & .698 \\
SAKI, rank-32 FP16 & 64 & .731 & .502 & .857 & .500 & .995 & .683 \\
Self-Indexing & 20 & .421 & .272 & .507 & .241 & .367 & .201 \\
\bottomrule
\end{tabular}
\end{table*}

\section{Implementation details}\label{app:C}

This appendix gives the implementation details behind \S\ref{sec:3.2}, \S\ref{sec:3.3} and \S\ref{sec:4}.

\textbf{C.1 Residual compression formulas.} Notation follows \S\ref{sec:3.2}. $k_p \in \mathbb{R}^d$ is the native index key at position $p$, with RoPE on its first $d_r$ dimensions. $r_p = k_p - m(p)$ is the residual of $k_p$.

\begin{itemize}
\item \emph{Shared component.} $m(p) = [R(p)\,v;\ \mu_n]$. Here $R(p)$ is the model's RoPE rotation at its deployed frequencies, including YaRN scaling (Appendix~\ref{app:B}.6). $v = \mathrm{E}[R(-p)\,k_p^{\mathrm{rot}}]$ is the calibration mean of the rotary dimensions before rotation. $\mu_n$ is the mean of the non-rotary dimensions.
\item \emph{Projection.} Let $\bar G$ be the shrunk, shift-averaged query metric of \S\ref{sec:3.2} ($\alpha = 0.5$ for Qwen and V4.1, 0 for GLM). Let $H = \bar G^{1/2}$, let $S_r$ be the covariance of the residuals, and let $U_D$ hold the top-$D$ eigenvectors of $H S_r H$. With $A = H U_D$ and $B = H^{-1} U_D$, $q\cdot k_p \approx q\cdot m(p) + (qB)\cdot(r_p A)$. We form $H^{-1}$ from the eigendecomposition of $\bar G$, with eigenvalues floored at $10^{-6}$ of the largest. The floor never applies at $\alpha = 0.5$, where $\bar G \succeq 0.5\,I$. Before quantization, the $\bar G$-weighted error of the centred residual equals $\sum_{i>D}\lambda_i(H S_r H)$, the minimum over rank-$D$ linear compressions in this metric (Appendix~\ref{app:B}.2). The compressed key does not store the residual mean, so centring only selects the directions of $U_D$.
\item \emph{Per-model fitting.} Qwen weights all index heads equally in $\Sigma_q$. V4.1 and GLM weight each head's queries by $\lvert w_h\rvert$. GLM also adds the rows that score highest under the native indexer to the residual covariance and its mean, with weight 4. GLM chooses each compressed key's NVFP4 scales by a search that minimizes the score error, not by rounding.
\item \emph{Scoring.} $\hat s_p = \sum_h w_h\,\mathrm{ReLU}\big(q_h\cdot m(p) + Q_4(q_h B)\cdot \hat z_p\big)$. Here $\hat z_p$ is the stored compressed key and $w_h$ are the model's head weights. $Q_4$ is the NVFP4 quantizer (E2M1 values, one E4M3 scale per 16 dimensions, a per-layer global scale). Twindex applies $Q_4$ once per step. Twindex evaluates the shared-component term by angle addition from a per-tile base rotation and a small per-offset cos/sin table.
\item \emph{Calibration (Qwen).} The fit uses five captures of four documents, 24,576 sampled blocks per capture per layer, and queries from five decode steps (100 query vectors over the four index heads). For each layer, the fit sets $A$, $B$, $v$, $\mu_n$ and the global scale, and D and M follow \S\ref{sec:adapt}.
\item \emph{Overlap check.} We compare the word 13-grams of the four complete calibration texts with LongBench v2 (all 503 contexts), the eight InfiniteBench tasks, RULER, BABILong with its PG19 background books \citep{ref78}, AIME 2025, GPQA-diamond \citep{ref55}, LongGenBench, LongProc \citep{ref57}, the throughput workload and the documents of the long-decoding tests. For LongGenBench and LongProc, the check covers each prompt and the 1M-token document placed before it. For every text and set, at most 6\ensuremath{\times}10\textsuperscript{\ensuremath{-}5} of the text's distinct 13-grams occur in the set. GLM's 18 calibration documents pass the same check, with at most 2.2\ensuremath{\times}10\textsuperscript{\ensuremath{-}4} of any document's 13-grams in any set. The one exception is Huckleberry Finn. It is part of Mark Twain's collected works, the context of one LongBench v2 item and the source of five throughput tails. The fit samples only the first 524,288 tokens of the Twain collection, and Huckleberry Finn lies outside this window. The 1M-token capture is the Go standard library.
\end{itemize}

\textbf{C.2 Compressed-key layouts.}

\begin{itemize}
\item \emph{Qwen.} Each compressed key holds D/2 bytes of 4-bit values and D/16 bytes of E4M3 scales, 36 B at D = 64. Twindex stores each compressed key as two segments: dimensions 0--63 and dimensions 64 to D\ensuremath{-}1. At the deployed D = 64 the second segment is empty. It serves wider variants. On the query side, Twindex quantizes the first D coordinates of qB to NVFP4 with a per-row global scale.
\item \emph{V4.1.} Layers 2 and 8 use D = 80, with dimensions 0--31 in FP4 and dimensions 32--79 in 2 bits (\{\ensuremath{\pm}0.5, \ensuremath{\pm}1.5\} times an E4M3 group scale). Each entry takes 33 B. Layers 14 and 20 use D = 64, with 16 FP4 and 48 2-bit dimensions, in 24 B per entry. Two converter warps expand the 2-bit part before the MMA.
\item \emph{GLM-5.2.} GLM-5.2 uses D = 64 on 3 layers, 48 on 8 layers and 32 on 10 layers. A compressed key takes 24.0 B per entry on average. We estimate that GLM's index hot buffer adds 6.9 B per entry at 992K tokens. This buffer holds native FP8 rows with their scales and tags. The total is 30.9 B per entry.
\end{itemize}

\textbf{C.3 Scan kernels.} For each tile, the Qwen scan issues two MMAs: one block-scaled NVFP4 MMA computes the compressed term and applies the E4M3 block scales inside the MMA, and one FP16 MMA adds the shared-component term into the same accumulator. The epilogue applies the per-head ReLU, the head weights and 1/\ensuremath{\surd}d, and writes subtile maxima for the radix select. In a one-GPU microbenchmark with 192 requests of 252,160 entries, a call reads the compressed keys and writes fp32 scores, 1.94 GB in all, in 388.6 \textmu{}s (5.0 TB/s). The V4.1 scans are raw-PTX kernels.

\textbf{C.4 Index hot buffers.} Qwen's index hot buffer is a ring of 8,192 native rows per request and layer. Each tag holds a CLOCK reference bit. Lookups go through a persistent hash table of 128 KiB per layer and request. Twindex rebuilds this table after 4,096 tombstones accumulate. A tombstone marks a deleted entry. V4.1's index hot buffers are 4-way set-associative with LRU replacement. They hold 4\ensuremath{\cdot}|\ensuremath{\mathcal{C}}| rows, where \ensuremath{\mathcal{C}} is the candidate set of \S\ref{sec:3.3}. At layer 20, the buffer holds four times the layer's candidate blocks, at 544 B each. Storing index rows invalidates the buffers.

\textbf{C.5 Compression paths.} On Qwen, Twindex writes the compressed key at every index-key store, in one launch that a CUDA graph can capture. On V4.1, Twindex compresses the newest entry on a side stream during decode. During prefill, Twindex compresses the closed entries of each chunk. V4.1 prefill also keeps an HBM mirror of the request's index rows (173 MiB per GPU), so prefill never reads host memory.

\section{Per-model configuration}\label{app:D}

Table~\ref{tab:1} lists how each model's index format, selection structure and costs set Twindex's configuration (\S\ref{sec:3.2}, \S\ref{sec:3.3} and \S\ref{sec:3.4}).

\begin{table*}[tp]
\centering
\footnotesize
\caption{The three models and Twindex's configuration for each.}\label{tab:1}
\begin{tabular}{@{}>{\raggedright\arraybackslash}p{\dimexpr 0.16\linewidth-2\tabcolsep\relax}>{\raggedright\arraybackslash}p{\dimexpr 0.26\linewidth-2\tabcolsep\relax}>{\raggedright\arraybackslash}p{\dimexpr 0.31\linewidth-2\tabcolsep\relax}>{\raggedright\arraybackslash}p{\dimexpr 0.27\linewidth-2\tabcolsep\relax}@{}}
\toprule
 & Qwen3.8-Flash-Next \citep{ref5} & DeepSeek V4.1-Flash \citep{ref6} & GLM-5.2 \citep{ref7} \\
\midrule
Native index entry & FP8, 132 B per 4-token block & FP4, 68 B per entry, one per 2 tokens (layers 2, 8, 14) or per token (layer 20) & FP8, 132 B per token \\
Indexer layers & 12 & 4 (layers 2, 8, 14, 20) & 21, other layers reuse their selections \\
Native top-k & 512 blocks & 512 per layer, layer 20 also picks 2,048 blocks for layers 24--36 & 2,048 \\
GPUs & 4 (TP4) & 4 (TP4) & 8 (TP8), index pages sharded \\
Twindex compressed key & D = 64, NVFP4, 36 B & FP4 + 2-bit, 33 B (layers 2, 8), 24 B (layers 14, 20) & D \ensuremath{\in} \{32, 48, 64\}, 24 B on average \\
Candidates & M = 1,280 blocks + set \ensuremath{\mathcal{A}} (\S\ref{sec:3.3}) & M = 3,072 / 1,792 / 3,584 (+128), layer 20: 6,144 blocks of 8, always including the 256 newest and 16 oldest & $M_l$ = 4,096--37,888, $M_l/8$ from each of 8 page classes \\
Index hot buffer & 8,192-row CLOCK ring & 4-way LRU, 4\ensuremath{\cdot}|\ensuremath{\mathcal{C}}| rows & 2,048--16,384 rows per layer \\
Index per request per GPU at 1M & 99.3 \ensuremath{\rightarrow} 30.7 MiB & 43.2 \ensuremath{\rightarrow} 21.6 MiB & 336 \ensuremath{\rightarrow} 79.1 MiB \\
Overlap of fetches & two-group schedule (both systems) & fetch and scan pipelined in chunks & indexer-layer swap-ins overlap the layers that reuse them \\
Query metric (\S\ref{sec:3.2}) & \ensuremath{\alpha} = 0.5, shifts \{0, \ensuremath{\pm}131,072, \ensuremath{\pm}262,144\} & \ensuremath{\alpha} = 0.5, same shifts & \ensuremath{\alpha} = 0, same shifts \\
Calibration & 5 captures of 4 docs (512K, one at 1M), 100 query vectors per layer & one 512K capture of each of the same 4 docs, 640 query vectors & one 512K capture of each of 18 other docs, 27,648 query vectors \\
\bottomrule
\end{tabular}
\end{table*}

\section{Additional results}\label{app:extra}

\begin{figure}[b]
\centering
\includegraphics[width=\columnwidth]{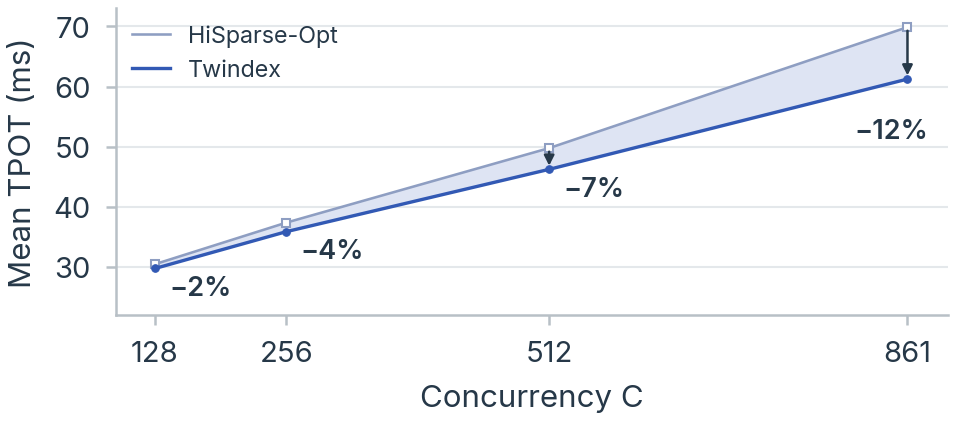}
\caption{Mean TPOT at equal concurrency (Qwen). Both systems run the same concurrencies in one session, and each label gives Twindex's change in TPOT.}\label{fig:eqc}
\end{figure}

This appendix gives the full-HBM throughput, the TPOT at equal concurrency (Figure~\ref{fig:eqc}) and the step-time breakdown (\S\ref{sec:5.3} and \S\ref{sec:lat}), the long-generation tests and the entries that Twindex misses (\S\ref{sec:5.4}), throughput with mixed lengths and many distinct prefixes and the concurrency limits by context length (\S\ref{sec:5.3}), the fitted cost model and the index-sharing runs (\S\ref{sec:lat} and \S\ref{sec:5.3}), and the sweeps of $\alpha$, D and M (\S\ref{sec:sens}).

\begin{figure}[tp]
\centering
\includegraphics[width=0.94\columnwidth]{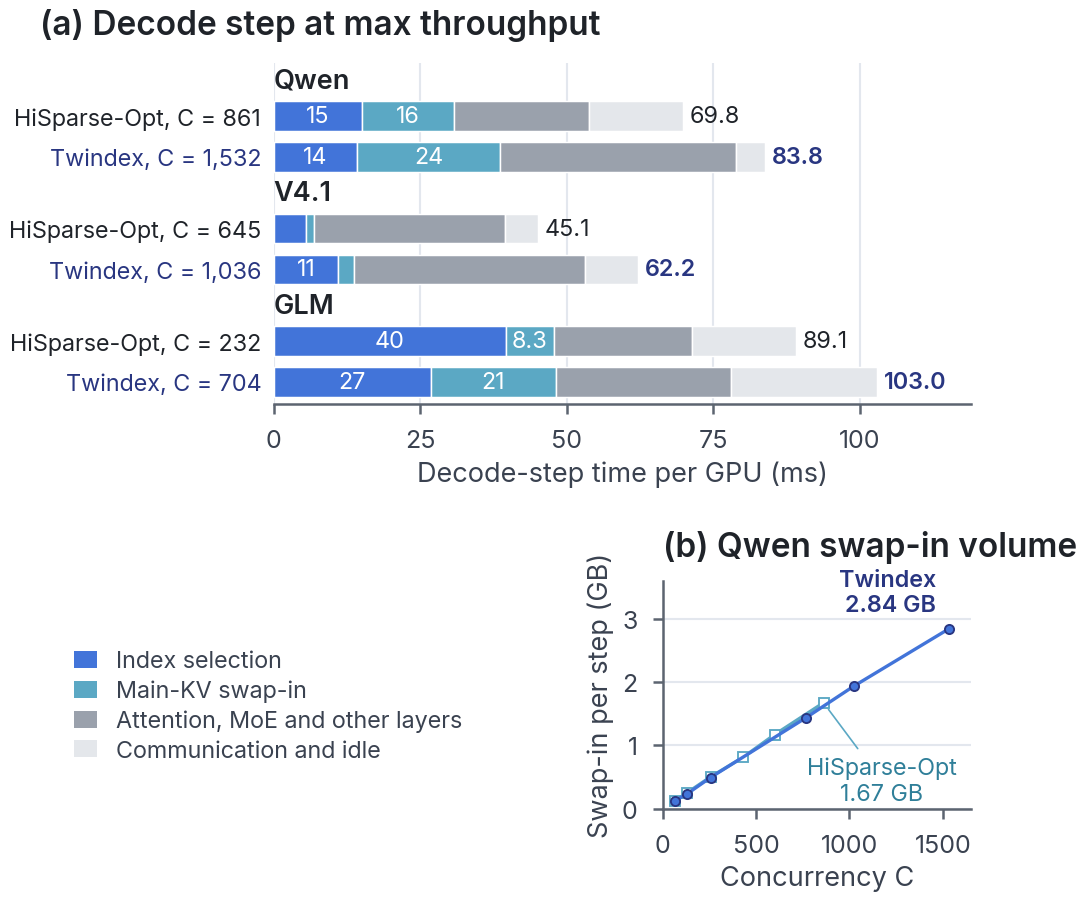}
\vspace{9pt}
\caption{Decode-step time at maximum throughput. (a) Decode-step time per GPU at each system's maximum-throughput C, from profiled steps. Twindex spends less time on index selection than HiSparse-Opt on Qwen and GLM, with 1.78\ensuremath{\times} and three times the requests, and more on main-KV swap-in and the other layers of the added requests. Kernels that overlap share their time. GLM's idle time includes profiler overhead. (b) Qwen's main-KV swap-in per step grows linearly with C in both systems.}\label{fig:brk}
\end{figure}

\begin{table*}[tp]
\centering
\footnotesize
\caption{Throughput at the full-HBM limit. 992K-token prompts on \GPU{} (TP4, GLM TP8), Q per serving instance. Qwen and GLM requests share three prefixes stored once in host memory, and V4.1 requests keep private copies. Each request keeps its own index in both systems. The PD rows give the steady-state decode throughput of a prefill/decode deployment (\S\ref{sec:5.3}). Qwen's PD row runs on another node, where the decode-node pair gives R = 1.463. V4.1's PD row also runs on another node.}\label{tab:2}
\begin{tabular}{@{}lrrrrrrr@{}}
\toprule
 & \multicolumn{3}{c}{HiSparse-Opt} & \multicolumn{3}{c}{Twindex} & \\
\cmidrule(lr){2-4}\cmidrule(lr){5-7}
Model & C & Step (ms) & Q (tokens/s) & C & Step (ms) & Q (tokens/s) & R \\
\midrule
Qwen & 861 & 69.82 & 12,332 & 1,532 & 86.39 & 17,734 & 1.438 \\
\quad PD, steady decode & 861 & 69.05 & 12,469 & 1,532 & 87.03 & 17,604 & 1.412 \\
V4.1 & 640 & 43.53 & 14,704 & 1,036 & 56.30 & 18,400 & 1.251 \\
\quad PD, steady decode & 630 & 43.54 & 14,471 & 1,036 & 58.45 & 17,725 & 1.225 \\
GLM & 232 & 79.73 & 2,910 & 704 & 95.25 & 7,391 & 2.540 \\
\bottomrule
\end{tabular}
\end{table*}

\begin{table*}[tp]
\centering
\footnotesize
\caption{Predicted and measured R at full HBM (\GPU{}, 1M context), from the fitted fixed cost F and per-request slope c of each system. Index share is the index's share of per-request HBM. Each pair is HiSparse-Opt, then Twindex.}\label{tab:18}
\begin{tabular}{@{}lcccccc@{}}
\toprule
Model & Index share & C & F (ms) & c (\textmu{}s) & Predicted R & Measured R \\
\midrule
Qwen3.8-Flash-Next & 63\% \ensuremath{\rightarrow} 35\% & 861 \ensuremath{\rightarrow} 1,532 & 23.1 / 24.3 & 53.9 / 41.3 & 1.41 & 1.438 \\
GLM-5.2 & 90\% \ensuremath{\rightarrow} 69\% & 232 \ensuremath{\rightarrow} 704 & 40.9 / 30.0 & 164.9 / 94.5 & 2.49 & 2.540 \\
\bottomrule
\end{tabular}
\end{table*}

\textbf{E.1 Long generations.} We test whether Twindex's selection drifts over 32,768-token outputs on Qwen after prefixes of about 1M tokens in three ways. First, Twindex's final set holds .943--.950 of the exact top-512 entries along its own 32,768-token trajectory. On V4.1 it holds .957--.982 at layers 2, 8 and 14 over eight documents with 983K-token prefixes, and no window of 512 steps falls below .947. Second, each system samples 32,768 tokens, and the unmodified model then reads each system's output in one prefill. At every output token, we measure the KL divergence from the unmodified model's next-token distribution to the generating system's, over the tokens in both top-64 lists, with the remaining probability pooled into one bucket. Over tokens 16K--32K it is .0047 for Twindex, .0056 for the baseline and .0062 for a second baseline run. Third, we fork greedy decoding from the document text at depths of 1K--16K tokens and count the share of the next 256 tokens that match the document's own continuation. Twindex's share differs from the baseline's by $-$.126 to $+$.014, within the $-$.133 to $+$.081 by which a second baseline run differs from the first. On V4.1 the shares differ by $-$.005 to $+$.025, within the second run's $-$.006 to $+$.038. Twindex is within .014 of the baseline on Qwen's AIME25 and LongGenBench.

\textbf{E.2 Step-time model.} Two properties of each model set R. The index share of per-request HBM sets how much Twindex raises C, and the per-request cost of a step sets how much of the gain remains once C is large. We model the step time as $E(C) = F + c\cdot C$, where F is the fixed cost of a step, such as reading the weights, and c is the time that each running request adds. Then $R = \frac{C_S}{C_B}\cdot\frac{E_B}{E_S} = \tau_B/\tau_S$, where S is Twindex, B is HiSparse-Opt and $\tau = F/C + c$ is one request's cost per step. This model predicts R = 1.41 on Qwen against 1.438 measured, and 2.49 on GLM against 2.540 (Table~\ref{tab:18}). Qwen gains from both terms: Twindex fits 1.78\ensuremath{\times} as many requests and lowers the slope from 53.9 to 41.3 \textmu{}s per request. V4.1, whose index is already FP4, gains from the larger batch. Repeated Qwen runs agree within 0.46\%.

\textbf{E.3 Missed entries.} The candidates miss native entries that hold 0.43\% of the native top-512's mass (\S\ref{sec:3.3}), and the entries that replace them win back 70\% of it, which leaves the net 0.13\%. The compressed scores rank entries close to the native order (correlation .947, Figure~\ref{fig:8}a), so the entries it misses rank low: 88.1\% of them lie at native ranks 257--512, which hold only 5.3\% of the mass.

\begin{figure*}[tp]
\centering
\includegraphics[width=\textwidth]{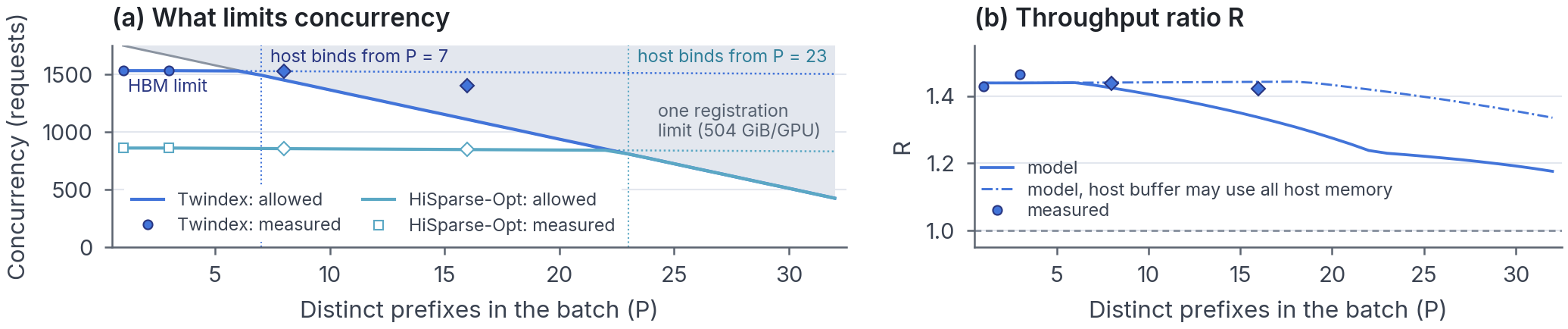}
\caption{Concurrency and R with P distinct prefixes (Qwen, 1M tokens). Twindex's gain holds with many distinct prefixes. (a) Each system's HBM limit (dotted), the 504 GiB limit of one pinned host registration (gray) and the concurrency each can run. Markers are measured, and the diamonds at P = 8 and 16 register the host arena in pieces. (b) R, measured and modeled. The dash-dot line lets the host buffer use all host memory.}\label{fig:11}
\end{figure*}

\begin{table}[tp]
\centering
\footnotesize
\caption{Recall by blend weight $\alpha$ on held-out documents. Plain PCA is $\alpha = 1$, and bold marks the deployed value. Recall on held-out documents: Qwen top-512 within 1,280 candidates at 128K--992K, V4.1 the same at 512K, GLM top-2,048 within half the served M at 992K.}\label{tab:alpha}
\setlength{\tabcolsep}{6pt}
\begin{tabular}{@{}lccc@{}}
\toprule
$\alpha$ & Qwen & V4.1 & GLM \\
\midrule
0 (query metric only) & .871 & .937 & \textbf{.978} \\
0.25 & .942 & .964 & -- \\
0.5 & \textbf{.949} & \textbf{.964} & .962 \\
1 (plain PCA) & .896 & .839 & .944 \\
\bottomrule
\end{tabular}
\end{table}

\begin{table}[tp]
\centering
\footnotesize
\caption{Recall by compressed width D and candidate count M on held-out documents. V4.1: native top-512 within M, one D on every layer, NVFP4 replay, six documents each scored by a fit that did not see it. GLM: native top-2,048 within M given as a multiple of the served $M_l$, 992K, six documents. The GLM mix is the served one, 24 B per entry.}\label{tab:sens}
\setlength{\tabcolsep}{4pt}
\begin{tabular}{@{}lccccc@{}}
\toprule
V4.1 & D = 48 & 64 & 80 & 96 & 128 \\
\midrule
Recall within 1,280 & .867 & .940 & .985 & .997 & 1.000 \\
M for recall .964 & 4,438 & 1,800 & 922 & 721 & 660 \\
\midrule
GLM, M / $M_l$ & 0.25 & 0.5 & 0.75 & 1 & 2 \\
\midrule
D = 32 & .861 & .962 & .986 & .993 & .999 \\
Served mix & .899 & .978 & .994 & .997 & 1.000 \\
D = 48 & .911 & .984 & .995 & .998 & 1.000 \\
D = 64 & .939 & .993 & .998 & .999 & 1.000 \\
\bottomrule
\end{tabular}
\end{table}

\textbf{E.4 Mixed lengths and tasks.} R is 1.35 on Qwen and 1.16 on V4.1 when 256K-, 512K- and 1M-token requests share a batch, and 1.05--1.22 on the four InfiniteBench \citep{ref50} tasks whose prompts have at least 256K tokens (Qwen).

\textbf{E.5 Many distinct prefixes.} Twindex's gain does not need many requests to share one prefix. V4.1 requests share no prefix state, and R is 1.25. On Qwen, with P distinct prefixes in the decode batch (P = 3 in the headline workload), R is 1.43 at P = 1, 1.46 at P = 3, 1.44 at P = 8 and 1.42 at P = 16, where each prefix serves about 88 Twindex requests (Figure~\ref{fig:11}). Each pair runs on one node, P = 1--8 on the node of Table~\ref{tab:2}'s PD row. At P = 8 and 16, the host arena is registered in pieces. At P = 8, HBM limits both systems, and Twindex runs 1,527 requests. At P = 16, the node's host memory limits Twindex to 1,403 requests.

\textbf{E.6 Context length.} On Qwen at 256K (Table~\ref{tab:3}), host memory limits Twindex to 1,943 requests, below its HBM limit, and at 512K HBM limits both systems. At 2M, the HBM left for requests holds 487 HiSparse-Opt requests, and our implementation caps Twindex at 1,024 requests (32-bit token index).

\textbf{E.7 Cost model.} Table~\ref{tab:18} gives the fitted fixed cost F and slope c of each system and the predicted R. On Qwen, we fit F and c on each system's step times below its full-HBM concurrency and predict the full-HBM pair. GLM is fitted as Qwen is.

\textbf{E.8 Index sharing.} On Qwen, both systems store each shared prefix's index once per GPU, keep each request's own tail index and run at equal C. When each request scans the shared index and both systems keep a main-KV buffer of 4,096 entries per request and layer, Twindex's step is 86.97 ms against 98.81 ms at C = 1,532 (R = 1.136) and 96.06 ms against 106.68 ms at C = 1,700 (R = 1.111). When both systems score the shared prefix in one batched pass and keep about 6,900 buffer entries, Twindex's step is 85.40 ms against 89.19 ms at C = 1,700 (R = 1.044), and a repeat pair on another node gives R = 1.042. Both systems then compute the same scores for the shared prefix, and Twindex compresses only each request's 16K-token private part. The footprint in \S\ref{sec:5.3} comes from replaying both systems' selections for 16 requests over 8,290 decode steps. It is the share of a request's context blocks that its selections touch at least once, averaged over requests and layers: 12.4--13.3\% at a 98\% prefix share and 18.6--19.1\% at a 47\% share.

\begin{figure}[tp]
\centering
\includegraphics[width=\columnwidth]{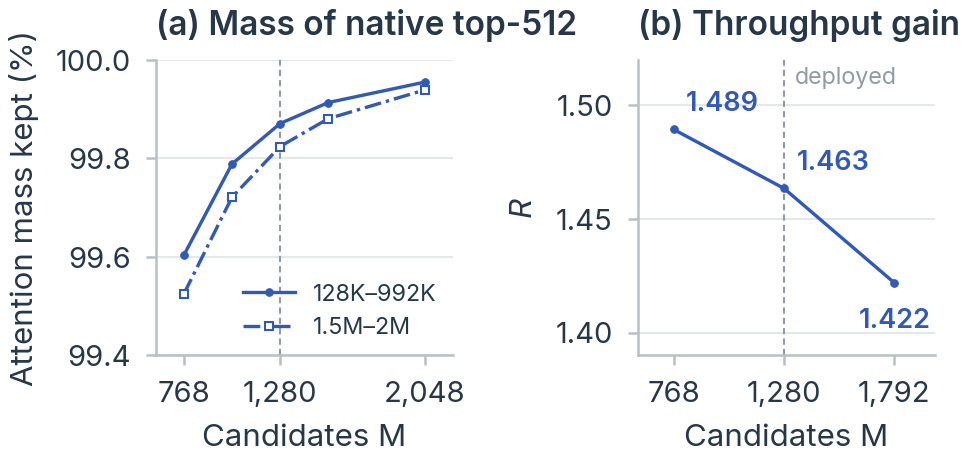}
\caption{Attention mass and R by candidate count M (Qwen). The dashed line marks the deployed M = 1,280. (a) Held-out documents. (b) R at 992K, one paired session.}\label{fig:14}
\end{figure}

\textbf{E.9 Candidates M.} Figure~\ref{fig:14} gives the attention mass and R by candidate count M (\S\ref{sec:sens}).

\label{docend}

\end{document}